\documentclass[11pt]{article}
\usepackage{amsmath}
\usepackage{graphicx}
\usepackage{enumerate}
\usepackage{natbib}
\usepackage{authblk}
\usepackage{url}  
\usepackage{comment}

\newcommand{\blind}{1}

\usepackage{xr-hyper}

\usepackage{amssymb,amsfonts,mathrsfs,amsmath,amsthm,mathtools,bm,dsfont, thmtools,bbm}\allowdisplaybreaks
\usepackage[dvipsnames]{xcolor}
\usepackage{array, graphicx, graphics,float,multirow,tabularx,tikz,pgfplots, longtable,threeparttable}
\usepackage{setspace}
\usepackage[colorlinks,citecolor=blue,urlcolor=blue, runcolor=blue, filecolor=cyan]{hyperref}
\usepackage{footnotebackref}
\usepackage{bibentry, natbib} 
\usepackage{paralist}  
\usepackage{appendix} 
\usepackage{mathrsfs}
\usepackage{enumitem}
\usepackage{authblk}
\usepackage{caption}
\usepackage{titlesec}
\usepackage{textcase,relsize}
\usepackage{datetime}
\usepackage{booktabs}

\declaretheoremstyle[notefont=\bfseries,notebraces={}{},%
    headpunct={},postheadspace=1em]{mystyle}
\declaretheorem[style=mystyle,numbered=no,name=Assumption]{asmp-hand}
\declaretheorem[style=mystyle,numbered=no,name=Condition]{cond-hand}
\declaretheorem[style=mystyle,numbered=no,name=Example]{exmp-hand}

	\def \calA {\mathcal{A}}		
			
	\def \calC {\mathcal{C}}		
			
\def \bbE {\mathbb{E}}	\def \calE {\mathcal{E}}		
	\def \calF {\mathcal{F}}		
	\def \calG {\mathcal{G}}		
	\def \calH {\mathcal{H}}		
	\def \calI {\mathcal{I}}

	\def \calM {\mathcal{M}}		
	\def \calN {\mathcal{N}}		
	\def \calO {\mathcal{O}}		
	\def \calP {\mathcal{P}}		
			
\def \bbR {\mathbb{R}}			
			
	\def \calT {\mathcal{T}}		
			
	\def \calV {\mathcal{V}}		
	\def \calW {\mathcal{W}}

	\def \calZ {\mathcal{Z}}

\def \Var {\mathrm{Var}}
\def \Cov {\mathrm{Cov}}

\def \conD {\overset{D}\longrightarrow}

\def \vect {\mathrm{vec}}

\def \diag {\mathrm{diag}}
\def \inv {\mathrm{inv}}

\def \infs {\mathrm{infs}}

\newcommand{\norm}[1]{\left\Vert#1\right\Vert}
\newcommand{\abs}[1]{\left\vert#1\right\vert}

\DeclareMathOperator*{\argmax}{arg\,max}
\DeclareMathOperator*{\argmin}{arg\,min}

\usepackage{subfig}

\numberwithin{equation}{section}

\theoremstyle{definition}
\newtheorem{remark}{Remark}

\newtheorem{assumption}{Assumption}[section]
\newtheorem*{assumption*}{Assumption}

\theoremstyle{plain}
\newtheorem{theorem}{Theorem}[section]
\newtheorem{proposition}[theorem]{Proposition}
\newtheorem{lemma}[theorem]{Lemma}

\def \conD {\overset{D}\longrightarrow}
\def \bbE {\mathbb{E}}	\def \calE {\mathcal{E}}		
\def \calP {\mathcal{P}}

\def \calW {\mathcal{W}}

\allowdisplaybreaks

\usepackage{algorithm,algpseudocode}

\makeatletter
\newenvironment{breakablealgorithm}
{
	\begin{center}
		\refstepcounter{algorithm}
		\hrule height.8pt depth0pt \kern2pt
		\renewcommand{\caption}[2][\relax]{
			{\raggedright\textbf{\fname@algorithm~\thealgorithm} ##2\par}%
			\ifx\relax##1\relax 
			\addcontentsline{loa}{algorithm}{\protect\numberline{\thealgorithm}##2}%
			\else 
			\addcontentsline{loa}{algorithm}{\protect\numberline{\thealgorithm}##1}%
			\fi
			\kern2pt\hrule\kern2pt
		}
	}{
		\kern2pt\hrule\relax
	\end{center}
}
\makeatother

\pgfplotsset{compat=1.16}
 
\def\spacingset#1{\renewcommand{\baselinestretch}%
	{#1}\small\normalsize} \spacingset{1}

\newdateformat{mydate}{\monthname[\THEMONTH] \THEYEAR}
\begin{document}

\def\spacingset#1{\renewcommand{\baselinestretch}%
{#1}\small\normalsize} \spacingset{1}

\onehalfspacing
\if1\blind
{
  \title{\bf Inference for Low-rank Completion without Sample Splitting with Application to Treatment Effect Estimation}
\author[1]{Jungjun Choi\thanks{Corresponding Author: \href{mailto:jc5805@columbia.edu}{jc5805@columbia.edu}}}
\author[2]{Hyukjun Kwon\thanks{\href{mailto:hk731@rutgers.edu}{hk731@rutgers.edu}}}
\author[3]{Yuan Liao\thanks{\href{mailto:yuan.liao@rutgers.edu}{yuan.liao@rutgers.edu}}}
\affil[1]{{\small Columbia University, 1255 Amsterdam Avenue New York, NY 10027, United States}}
\affil[2,3]{{\small Rutgers University, 75 Hamilton St, New Brunswick, NJ 08901, United States}}
  \maketitle
} \fi

\spacingset{2}
\if0\blind
{
  \bigskip
  \bigskip
  \bigskip
  \begin{center}
    {\LARGE\bf Inference for Low-rank Completion without Sample Splitting with Application to Treatment Effect Estimation}
\end{center}
  \medskip
} \fi

\spacingset{1.3} 
\begin{abstract}
This paper studies the inferential theory for estimating  low-rank matrices. It also provides an inference method for the average treatment effect as an application. We show that the least square estimation of eigenvectors following the nuclear norm penalization attains the asymptotic normality. The key contribution of our method is that it does not require sample splitting.  In addition, this paper allows dependent observation patterns and heterogeneous observation probabilities. Empirically, we apply the proposed procedure to estimating  the impact of the presidential vote on allocating the U.S. federal budget to the states.
 
\end{abstract}

\noindent%

\noindent%

\spacingset{1.5} 

\setlength{\abovedisplayskip}{8pt}
\setlength{\belowdisplayskip}{8pt}
\setlength\intextsep{8pt}
\setlength{\abovecaptionskip}{4pt}

\section{Introduction } \label{sec:intro}
	
The task of imputing the missing entries of a partially observed matrix, often dubbed as \textit{matrix completion}, is widely applicable in various areas. In addition to the well-known application to recommendation systems (e.g., the Netflix problem), this problem is applied in a diverse array of science and engineering such as collaborative filtering, system identification, social network recovery, and causal inference.  
	
 In this paper, we focus on the following approximate low-rank model with a factor structure: 
	\begin{align}\label{eq:starting}
		Y = M + \calE \approx \beta F' + \calE,
	\end{align}
	where $Y$ is an $N \times T$ data matrix which is subject to missing, $M$ is a latent matrix of interest, and $\calE$ represents a noise contamination. Importantly, $M$ is assumed to be an approximate low-rank matrix having an approximate factor structure $M \approx \beta F'$, where $\beta$ is factor loadings and $F$ is latent factors. In addition, we allow some entries of $Y$ to be unobserved by defining an indicator $\omega_{it}$, which equals one if the $(i,t)$ element of $Y $ is observed, and zero otherwise. In this practical setting, we provide the inferential theory for each entry of $M$, regardless of whether its corresponding entry in $Y$ is observed or not.
	
	One of the widely used methods for the low-rank matrix completion is the nuclear norm penalization and it has been intensively studied in the last decade. \cite{candes2009exact}, \cite{candes2010matrix}, \cite{koltchinskii:2011}, \cite{negahban2012restricted}, and \cite{chen2020noisy} provide statistical rates of convergence for the nuclear norm penalized estimator and a branch of studies including \cite{beck:2009}, \cite{cai:2010}, \cite{mazumder:2010}, \cite{ma:2011}, and \cite{parikh2014proximal} provide algorithms to compute the nuclear norm penalized estimator. However, research on inference is still limited. This is because the shrinkage bias caused by the penalization, as well as the lack of the closed-form expression of the estimator, hinders the distributional characterization of the estimator.

	We contribute to the literature by providing an inferential theory of the low-rank estimation without sample splitting. Our estimation procedure consists of the following main steps: 
	\begin{enumerate}[itemsep=-2mm]
	    \item Using the full sample of observed $Y$, compute the nuclear norm penalized estimator $\widetilde{M}$ and use the left singular vectors of $\widetilde{M}$ as the initial estimator for $\beta$.
	    \item To estimate $F$, regress the  observed $Y$ onto the initial estimator for $\beta$.
	    \item To re-estimate $\beta$, regress the  observed $Y$ on the estimator for $F$.
	    \item The product of the estimators in Steps 2 and 3 is the final estimator for $M$. 
	\end{enumerate}
	Note that steps 2-3 are only conducted once without further iterations. 
	
	An important contribution is that we do not rely on the sample splitting to make inference, but simply use the full (observed) sample in every step of our procedure. There are at least three advantages to avoid sample splitting. First, the resulting estimator using sample splitting is unstable and random even conditioning on the data. Second, sample splitting requires relatively large $T$ in practice, because it practically works with only $T/2$ observations. This is demanding in applied micro applications when $T$ is just a few decades. In the simulation study, we show that the performance of the estimator using sample splitting is worse than that of the estimator without sample splitting when $T$ is relatively small. Lastly, sample splitting increases computational costs in multiple tests because for each target time `$t$', we need to use different sample splitting.

	Technically, we apply a new approach to showing the negligibility of the potential bias terms, by making use of a hypothetically defined \textit{auxiliary  leave-one-out} (ALOO) estimator. We emphasize the word ``auxiliary" because it is only introduced in the technical argument, but \textit{not} implemented in the estimation. So it is a hypothetical estimator, which is to be shown that it is 
	
	i) asymptotically equivalent to the initial estimator for $\beta$ in Step 1 and 
	
	ii) independent of the sample used in the least squares estimation, namely, the sample in period $t$. Using the ALOO estimator, we can separate out the part in the initial estimator for $\beta$, which is correlated with the sample in period $t$. Once we separate out the correlated part, we can enjoy a similar effect to the sample splitting. And we show the separated correlated part is sufficiently small. Importantly, the leave-one-out estimator only appears in the proof as an auxiliary point of the initial estimator for $\beta$, so we do not need to compute it in the estimation procedure, which allows us to remove the sample splitting step without implementing any additional steps. 
	
		Empirically, we apply the proposed procedure to making inference for the impact of the presidential vote on allocating the U.S. federal budget to the states. 
	We find the states that supported the incumbent president in past presidential elections tend to receive more federal funds and this tendency is stronger for the loyal states than the swing states. In addition,  this tendency is stronger after the 1980s.

	\subsection{Relation to the literature}
Very recently, some studies proposed the ways of achieving unbiased estimation for the inference of the nuclear norm penalized estimator.   \cite{chernozhukov2019inference, chernozhukov2021inference} propose a  two-step least square procedure with sample splitting, which estimates the factors and loadings successively using the least square estimations. As we discussed earlier, sampling splitting comes with several undesirable costs.

	The idea of the ALOO estimator has been employed in other recent works such as \cite{ma2019implicit,chen:2019inference,chen2020nonconvex,chen2020noisy,yan2021inference} as well. Among them, in particular,
	\cite{chen:2019inference} pioneered using this idea to convex relaxation of low-rank inference. This paper has some important contributions compared to \cite{chen:2019inference}.
	\begin{enumerate}
	    \item We consider a general nonparametric panel model which is an approximate low-rank model rather than an exact low-rank model.
	    \item This paper accommodates more general data-observation patterns: the heterogeneous observational probabilities and the correlated observation patterns by assuming the cluster structure and allowing dependence within a cluster.
	    \item The inferential theory for the average treatment effect estimation is provided as an application.
	    \item We formally address a technical issue concerning the ALOO   estimator. The  ALOO 
 estimator is to be (hypothetically) calculated by using the gradient descent iteration from the leave-one-out problem, which rules out, for example, samples in period $t$. This exclusion is designed to guarantee the independence between the leave-one-out estimator and the period $t$ sample. However, due to the non-convexity of the loss functions, the gradient descent iteration must stop where the gradient of the loss function is sufficiently ``small.'' If this stopping point depends on the sample in period $t$, as in \cite{chen:2019inference} who derive the stopping point from the problem using the full sample, the leave-one-out estimator using this stopping point may not be truly independent of the sample in period $t$. This dependence frustrates the analysis of the bounds regarding the leave-one-out estimator. We provide two solutions for this potential dependence issue to be detailed in the paper. 
	    \item Our method does not have an explicit debias step, but is based on refitting least squares. While we do not claim that this estimator is advantageous over the explicit debiasing method, we view our estimator as the natural extension of ``post model selection methods'' to the low rank framework.
	\end{enumerate}

 Other related works on inference include \cite{xia2021statistical}, \cite{xiong2020large}, and \cite{jin2021factor}. We compare these methods with ours in simulations.

Lastly, a comparison with other literature that takes advantage of a low-rank model to estimate the treatment effect would be helpful. The close connection between low-rank completion and treatment effect estimation
was first made formal by \cite{athey2021matrix} who showed that the nuclear norm regularization can be useful for causal panel data by presenting the convergence rate of the estimator.  
Another   line of research proposes inferential theories under weaker assumptions on the treatment assignment with other restrictions. \cite{farias2021learning} allow the assignment of the treatment that can depend on historical observations while focusing on the estimation of the average treatment effect. \cite{agarwal2021causal} and \cite{bai2021matrix} consider the case where the assignment is not random but has a certain block structure that often occurs in causal panel data.\footnote{In \cite{agarwal2021causal}, a certain submatrix for estimation has a block structure.} In addition, \cite{arkhangelsky2021synthetic} propose an estimator that is more robust than the conventional difference-in-differences and synthetic control methods by using a low-rank fixed effect model with the homogeneous treatment effect assumption.

	This paper is organized as follows. Section \ref{sec:modelandestimation} provides the model and the estimation procedure as well as our strategy for achieving the unbiased estimation. Section \ref{sec:result} gives the asymptotic results of our estimator. Section \ref{sec:treatment} provides the inferential theory for the average treatment effect estimator as an application. Section \ref{sec:empirical} presents an empirical study about the impact of the president on allocating the U.S. federal budget to the states to illustrate the use of our inferential theory. Section \ref{sec:sim} includes the simulation studies. Section \ref{sec:conclusion} concludes.
	
	There are a few words on our notation. For any matrix $A$, we use $\norm{A}_F$, $\norm{A}$, and $\norm{A}_*$ to denote the Frobenius norm, operator norm, and nuclear norm respectively. $\norm{A}_{2,\infty}$ denotes the largest $l_2$ norm of all rows of a matrix $A$. $\vect(A)$ is the vector constructed by stacking the columns of the matrix $A$ in order. Also, $\psi_r(A)$ is $r$th largest singular value of $A$. $\psi_{\max}(A)$ and $\psi_{\min}(A)$ are the largest and the smallest nonzero singular value of A. For any vector $B$, $\diag(B)$ is the diagonal matrix whose diagonal entries are $B$. $a\asymp b$ means $a/b $ and $b/a$ are $O_P(1)$.

	\section{Model and Estimation} \label{sec:modelandestimation}
	We consider the following nonparametric panel model subject to missing data problem:
    \[
	y_{it}  = h_t\left( \zeta_{i}  \right) + \varepsilon_{it},
	\]
	where $y_{it}$ is the scalar outcome for a unit $i$ in a period $t$, $h_t(\cdot)$ is a time-varying nonparametric function, $ \zeta_{i}$ is a unit-specific latent state variable, $\varepsilon_{it}$ is the noise, and $\omega_{it} =1{\{y_{it}  \text{ is observed}\}}$.  Here, $\{h_t( \cdot ), \zeta_{i},  \varepsilon_{it}\}$ are unobservable. In the model, the (latent) unit states $ \zeta_{i}$ have a time-varying effect on the outcome variable through $h_t(\cdot)$. This model can be written in \eqref{eq:starting} using the sieve representation. Suppose the function $h_t(\cdot)$ has the following sieve approximation:
	\[
	h_t( \zeta_{i}) = \sum_{r=1}^{K} \kappa_{t,r}  \phi_{r}( \zeta_{i}) + M^{R}_{it} = \beta_{i}^{\prime}F_t   + M^{R}_{it} = M^{\star}_{it}  + M^{R}_{it} ,
	\]
	where $\beta_{i} = (\phi_{1}( \zeta_{i}),\dots, \phi_{K}( \zeta_{i}))^{\prime}$ and $F_t = (\kappa_{t,1},\dots,\kappa_{t,K})^{\prime}$. Here, $M_{it}^{R}$ is the sieve approximation error and, for all $1\leq r \leq K$, $\phi_{r}( \zeta_{i})$ is the sieve transformation of $\zeta_i$ using the basis function $\phi_{r}( \cdot)$ and $\kappa_{t,r}$ is the sieve coefficient. Then, 
 $$
 M=[M_{it}]_{N\times T},\quad M_{it}= h_t(\zeta_{i})
 $$
  can be successfully represented as the approximate factor structure.  
	
	In matrix form, we can represent the model as
	\begin{align}\label{eq:matrixform}
		Y = M + \calE = M^{\star} + M^{R} +\calE =  \beta F' + M^{R} + \calE,
	\end{align}
	where we denote $Y = [y_{it}]_{N\times T}$, $M = [M_{it}]_{N\times T}$, $M^{\star} = [M^{\star}_{it}]_{N\times T}$, $M^{R} = [M^{R}_{it}]_{N\times T}$, $\beta = [\beta_{1}, \ldots , \beta_{N} ]'$, $F = [F_1, \ldots , F_T ]'$, and $\calE = [\varepsilon_{it}]_{N\times T}$. Note that $Y$ is an incomplete matrix that has missing components. 
	
	Let $\mathcal{M} \coloneqq (\beta, F,M^{R})$ be the triplet of random matrices that compose $M$. In the paper, we allow the heterogeneous observation probability, i.e., $P(\omega_{it}=1)=p_i$ and denote $\Pi = \diag(p_1,\dots,p_N)$. Here, we shall assume the sieve dimension $K$ is pre-specified by researchers and propose some data-driven ways of choosing $K$ in Section \ref{sec:choosingK} of Appendix.

	\subsection{Nuclear norm penalized estimation with inverse probability weighting} \label{subsec:obfunction}

	To accommodate the heterogeneous observation probability, this paper uses the inverse probability weighting scheme, referred to as inverse propensity scoring (IPS) or inverse probability weighting in causal inference literature (e.g., \cite{imbens:2015}, \cite{little2019statistical}, \cite{schnabel2016recommendations}), in addition to the nuclear norm penalization:
	\begin{align}\label{eq:convexob}
		\widetilde{M}\coloneqq  \argmin_{A \in \mathbb{R}^{N \times T}} \frac{1}{2}\|\widehat{\Pi}^{-\frac{1}{2}}\Omega \circ \left( A - Y \right)\|_F^2 + \lambda\|A\|_*
	\end{align}
	where $\widehat{\Pi} = \diag(\widehat{p}_1, \dots, \widehat{p}_N)$, and $\widehat{p}_i = \frac{1}{T}\sum_{t=1}^{T}\omega_{it}$ for each $i\leq N$, $\Omega = [\omega_{it}]_{ N\times T}$ and $\circ$ denotes the Hadamard product. As noted in \cite{ma2019missing}, this inverse probability weighting debiases the objective function itself. If there is heterogeneity in the observation probability, $\|\Pi^{-\frac{1}{2}}\Omega \circ\left( A - Y \right)\|_F^2$ is an unbiased estimate of $\norm{A-Y}_F^2$, which we would use if there is no missing entry, in the sense that $\bbE_{\Omega}[\|\Pi^{-\frac{1}{2}}\Omega \circ\left( A - Y \right)\|_F^2]=\norm{A-Y}_F^2$, while $\norm{\Omega \circ(A-Y)}_F^2$ is biased.

	\subsection{Estimation procedure}\label{subsec:formalestimationalgorithm}
	
	Although the inverse probability weighting enhances the estimation quality, the weighting alone cannot guarantee the asymptotic normality of the estimator because of the shrinkage bias. To achieve the unbiased estimation having the asymptotic normality, we run the two-step least squares procedure. As noted previously, our estimation does not have the sample splitting steps.	Our estimation algorithm is as follows:
	
	\begin{breakablealgorithm}
		\caption{\small  Constructing the estimator for $M$.}
		\label{alg:estimation}
		\begin{algorithmic}
			\noindent \textbf{Step 1} Compute the initial estimator $\widetilde{M}$ using the nuclear norm penalization. \\
			\textbf{Step 2} Let $\widetilde{\beta}$ be $N \times K$ matrix whose columns are $\sqrt{N}$ times the top $K$ left singular vectors of $\widetilde{M}$.\\
			\textbf{Step 3} For each $t \leq T$, run OLS to get $\widehat{F}_t = \left( \sum_{j=1}^{N}\omega_{jt}\widetilde{\beta}_{j} \widetilde{\beta}_{j}^{\prime}   \right)^{-1}\sum_{j=1}^{N}\omega_{jt} \widetilde{\beta}_{j} y_{jt}$.\\
			\textbf{Step 4} For each $i \leq N$, run OLS to get $\widehat{\beta}_{i}=\left( \sum_{s=1}^{T} \omega_{is}\widehat{F}_s\widehat{F}_s^{\prime} \right)^{-1} \sum_{s=1}^{T}\omega_{is}\widehat{F}_s y_{is}$.\\
			\textbf{Step 5} The final estimator $\widehat{M}_{it}$ is $\widehat{\beta}_{i}^{\prime}\widehat{F}_t $ for all $(i,t)$.
		\end{algorithmic}
	\end{breakablealgorithm}

	After deriving the initial estimator of loadings from the nuclear norm penalized estimator $\widetilde{M}$, we estimate latent factors and loadings using the two-step least squares procedure. The final estimator of $M$ is then the product of the estimates for latent factors and loadings. 
	
	\subsection{A general discussion of the main idea}\label{subsec:debiasstrategys}

		It is well-known that the nuclear-norm penalized estimator $\widetilde{M}$, like other penalized estimators, is subject to shrinkage bias which complicates statistical inference. To resolve this problem, we use the two-step least squares procedure, i.e., Steps 3 and 4 in Algorithm \ref{alg:estimation}. In showing the asymptotic normality of the resulting estimator $\widehat{M}$, a key challenge is to show the following term is asymptotically negligible:
		\[
		    R_t = \frac{1}{\sqrt{N}} \sum_{j=1}^N \omega_{jt}\varepsilon_{jt}(\widetilde{\beta}_j - H_1'\beta_j) 
		\]
		where $H_1$ is some rotation matrix.\footnote{Another term $\frac{1}{\sqrt{N}} \sum_{j=1}^N (\omega_{jt}-p_j)\beta_j F_t'H_1^{\prime -1}(\widetilde{\beta}_j - H_1'\beta_j)$ is also to be shown negligible, but the argument is similar to that of $R_t$.} This term represents the effect of the bias of the nuclear-norm penalization since $\widetilde{\beta}_j$ is derived from the nuclear-norm penalized estimator. \cite{chernozhukov2019inference,chernozhukov2021inference} resort to sample splitting to show the asymptotic negligibility of $R_t.$

\subsubsection{The auxiliary leave-one-out method}	\label{subsubsec:loo}
 
Motivated by \cite{chen2020noisy}, we show the asymptotic negligibility of $R_t$ without sample splitting by using two hypothetical estimators which are asymptotically equivalent to the nuclear norm penalized estimator $\widetilde{\beta}$. Namely, we consider a hypothetical non-convex iteration procedure for the low-rank regularization, where singular vectors are iteratively solved as the solution and show that this procedure can be formulated as the following two problems: 
	\begin{align}
	\label{eq:nonconvexloss} L^{\mathrm{full}}(B,F)   
	&=  \frac{1}{2}\|\Pi^{-\frac{1}{2}}\Omega \circ\left( B F' - Y \right)\|_F^2 + \frac{\lambda}{2}\|B\|_F^2 + \frac{\lambda}{2}\|F\|_F^2 \nonumber\\
	&=\frac{1}{2}\|\Pi^{-\frac{1}{2}}\Omega \circ\left( B F' - Y \right)\|_{F,(-t)}^2+\frac{1}{2}\|\Pi^{-\frac{1}{2}}\Omega \circ\left( B F' - Y \right)\|_{F,t}^2 + \frac{\lambda}{2}\|B\|_F^2 + \frac{\lambda}{2}\|F\|_F^2 \\
		\label{eq:leaveoneoutloss}L^{(-t)}(B, F)  &=\frac{1}{2}\|\Pi^{-\frac{1}{2}}\Omega \circ\left( B F' - Y \right)\|_{F,(-t)}^2+\frac{1}{2}\| B F'-M^{\star}\|_{F,t}^2 + \frac{\lambda}{2}\|B \|_F^2 + \frac{\lambda}{2}\|F\|_F^2.
\end{align}
Here, $\| \cdot \|_{F,(-t)}$ denotes the Frobenius norm which is computed ignoring $t$-th column and $\| \cdot \|_{F,t}$ is the Frobenius norm of only $t$-th column. Note that the only difference between \eqref{eq:nonconvexloss} and \eqref{eq:leaveoneoutloss} is that the $t$-th column of the goodness of fit part in \eqref{eq:nonconvexloss} is replaced by its conditional expectation in \eqref{eq:leaveoneoutloss}. So, $\{\omega_{jt}, \varepsilon_{jt}\}_{j \leq N}$ is excluded from the problem \eqref{eq:leaveoneoutloss}. 

We emphasize that (i) both problems defined above are non-convex; (ii) both problems are ``auxiliary", meaning that they are introduced only for proofs, not actually implemented. (iii) Optimizing $L^{(-t)}(B, F) $ is an  auxiliary leave-one-out (ALOO)  problem, leading to the ALOO estimator $\breve{\beta}^{(-t)} $ to be discusssed below.

Because of the non-convexity, 
both hypothetical problems should be computed iteratively until the gradients of the non-convex loss functions become ``sufficiently small.'' However, the gradients do not monotonically decrease as iteration proceeds since the problem is non-convex. So, one cannot let it iterate until convergence is reached, but has to stop at the point where the gradient is small enough. The choice of this ``stoping point'' is crucial in the analysis of the residual terms. \cite{chen:2019inference} define the stopping point using the full sample problem \eqref{eq:nonconvexloss}, which potentially causes dependence problem of the leave-one-out estimators. We propose two approaches of addressing this issue.
\begin{description}
    \item[Approach I] First, we derive the stopping point from the leave-one-out problem \eqref{eq:leaveoneoutloss}. Let $B^{\mathrm{full},\tau}$ and $B^{(-t),\tau}$ be $\tau$-th iterates of the gradient decent for \eqref{eq:nonconvexloss} and \eqref{eq:leaveoneoutloss}, respectively. Fix $t$ of interest and suppose we iterate both problems $\tau_t$ times, where $\tau_t$ depends on $t$. Define the ``solutions'' at $\tau_t$-th iterations: 
$$
  \breve{\beta}^{\mathrm{full},t}=B^{\mathrm{full},\tau_t} \quad \text{and} \quad \breve{\beta}^{(-t)}=B^{(-t),\tau_t}. 
$$
\noindent Hence, they share the same stopping point $\tau_t$. Noticeably, although $\breve{\beta}^{\mathrm{full},t}$ is a solution for the full sample problem \eqref{eq:nonconvexloss}, it depends on $t$ through $\tau_t.$ In this first approach, we derive the stopping point from the ALPOO problem \eqref{eq:leaveoneoutloss}. Hence, it ensures that the estimator $\breve{\beta}^{(-t)}$ using this stopping point is independent of the $t$-th period sample, $\{\omega_{jt}, \varepsilon_{jt}\}_{j \leq N}$.
This introduces nontrivial technical challenges. Namely, $\tau_t$, being derived from the problem $L^{(-t)}(B,F)$, depends on $t$, so the ``full-problem" solution $\breve{\beta}^{\mathrm{full},t}$  would therefore also depend on $t$. We derive the uniform convergence of both $\breve{\beta}^{\mathrm{full},t}$  and $\breve{\beta}^{(-t)}$ uniformly in $t=1,...,T.$

Being equipped with these two auxiliary non-convex estimators, we can bound $R_t$ in the following scheme: 
\begin{enumerate}
    \item First, decompose $R_t$ into two terms: 
	\begin{align}
		    R_t &= \frac{1}{\sqrt{N}} \sum_{j=1}^N \omega_{jt}\varepsilon_{jt}(\widetilde{\beta}_j - H_1'\beta_j) \nonumber\\
 &=\underbrace{\frac{1}{\sqrt{N}} \sum_{j=1}^N \omega_{jt}\varepsilon_{jt}(\widetilde{\beta}_j - \breve{\beta}^{(-t)}_j)}_{\coloneqq a} + \underbrace{\frac{1}{\sqrt{N}} \sum_{j=1}^N \omega_{jt}\varepsilon_{jt}(\breve{\beta}^{(-t)}_j - H_1'\beta_j).}_{\coloneqq b} \label{eq:Rtdecompose}
		\end{align} 
    \item $\max_t\norm{b}=o_P(1)$ can be shown  relatively easily due to the genuine independence between $\breve{\beta}^{(-t)}$ and $\{\omega_{jt}\varepsilon_{jt}\}_{j \leq N}$, which is along the same line as sample splitting.  Importantly,  it is crutial to require that $\tau_t$ should not depend on observations of time  $t$. So the stopping time should be defined carefully, which is one of the main technical contributions of the paper.

    \item In addition, $\max_t\norm{a}=o_P(1)$ comes from the following two rationales:
    $$
    a=\frac{1}{\sqrt{N}} \sum_{j=1}^N \omega_{jt}\varepsilon_{jt}(\widetilde{\beta}_j -\breve{\beta}^{\mathrm{full},t}_j )  + \frac{1}{\sqrt{N}} \sum_{j=1}^N \omega_{jt}\varepsilon_{jt}( \breve{\beta}^{\mathrm{full},t}_j-    \breve{\beta}^{(-t)}_j).
    $$
    \begin{enumerate}
        \item $\breve{\beta}^{\mathrm{full},t} \approx \breve{\beta}^{(-t)}$\\ 
        Their loss functions \eqref{eq:nonconvexloss} and \eqref{eq:leaveoneoutloss} are very similar and they share the same stopping point $\tau_t$. Therefore, $ \max_t \| \breve{\beta}^{\mathrm{full},t} - \breve{\beta}^{(-t)} \|$ is sufficiently small. Following the guidance of \cite{chen2020noisy}, we apply  the mathematical induction.
        \item $\widetilde{\beta} \approx \breve{\beta}^{\mathrm{full},t}$\\
   Note that $\breve{\beta}^{\mathrm{full},t}$ is derived from the non-convex problem \eqref{eq:nonconvexloss} and $\widetilde{\beta}$ comes from the nuclear norm penalization \eqref{eq:convexob}. Although the loss functions \eqref{eq:convexob} and \eqref{eq:nonconvexloss} are seemingly distinct, their penalty terms are closely related in the sense that
        \[
            \|A\|_* = \inf_{B \in \mathbb{R}^{N \times K}, F \in \mathbb{R}^{T \times K} : B F'=A} \Big\{\frac{1}{2}\norm{B}_F^2+\frac{1}{2}\norm{F}_F^2 \Big\}.
        \]
        Hence, $\max_t \|\widetilde{\beta} - \breve{\beta}^{\mathrm{full},t}\|$ is sufficiently small. A technical innovation is that $\breve{\beta}^{\mathrm{full},t}$ depends on $t$ so the uniformity is crucially relevant.
    \end{enumerate}
    Hence, we have $\max_t\|R_t\| = o_P(1)$.
\end{enumerate}
 
    \item[Approach II] Alternatively, we can follow the definition of the stopping point in \cite{chen:2019inference}, which uses the full sample. And then, we correct their proof by showing that, although the leave-one-out estimator is not independent of the sample data in period $t$, we can still obtain a uniform bound over iterations. Denote the stopping point from \cite{chen:2019inference} as $\tau^*$. In lieu of $(B^{\mathrm{full},\tau_t}, B^{(-t), \tau_t})$, we use $(B^{\mathrm{full},\tau^*}, B^{(-t),\tau^*})$ as the solutions for \eqref{eq:nonconvexloss} and \eqref{eq:leaveoneoutloss}, respectively. 
    
    Recall the decomposition \eqref{eq:Rtdecompose}. The analysis of term $a$ is analogous to the previous case. Regarding term $b$, we highlight that $\breve{\beta}^{(-t)}$, which is $B^{(-t),\tau^*}$, is not independence from the sample in period $t$, i.e., $\{\omega_{jt}, \varepsilon_{jt}\}_{j \leq N}$, since the stopping point $\tau^*$ depends on it. We will provide a uniform bound over iteration $\tau$ and period $t$ for term $b:$
\begin{align*}
	\max_t\norm{b} &= \max_t\norm{\frac{1}{\sqrt{N}} \sum_{j=1}^N \omega_{jt}\varepsilon_{jt}(\breve{\beta}^{(-t)}_j - H_1'\beta_j)} =\max_t \norm{\frac{1}{\sqrt{N}} \sum_{j=1}^N \omega_{jt}\varepsilon_{jt}(B^{(-t),\tau^*}_j - H_1'\beta_j)}\\
	& \leq \max_t\max_{\tau} \norm{\frac{1}{\sqrt{N}} \sum_{j=1}^N \omega_{jt}\varepsilon_{jt}(B^{(-t),\tau}_j - H_1'\beta_j)} = o_P(1).
		\end{align*}

\end{description}

In either way, we can successfully show the negligibility of $R_t$ uniformly in $t$ without resorting to sample splitting. We highlight that the first approach is more natural in the sense that it automatically ensures the independence that we need for term $b$. Our first approach, while technically more involved, is potentially more applicable to general machine learning inferences that rely on auxiliary leave-one-out estimators, because of the natural independence. In contrast, it is unclear whether the second approach is still applicable in other cases.

\subsubsection{Why is the auxiliary leave-one-out problem defined in this way?}

It is natural to ask why would not we define the ALOO estimator  more naturally as the  original estimator $\widetilde\beta$, but simply dropping the 
$t$ th column from the data matrix in the optimization?     One of the key differences   between  $L^{(-t)}(B, F)$ in (\ref{eq:leaveoneoutloss}) and the ``more natural dropping-$t$" loss,  is that the $t$ th column in   the least squares part of  $L^{(-t)}(B, F)$ is  not  simply dropped, but is replaced by its expectation:
$$
\mathbb E  \|\Pi^{-\frac{1}{2}}\Omega \circ\left( B F' - Y \right)\|_{F,t}^2  =  \| BF'-M^{\star}\|_{F,t}^2 + C
$$
where the constant $C$ does not depend on $(B,F)$. 
The reason for defining the ALOO loss function in this way  is to gain ``hypothetical efficiency", so that the ALOO estimator would be closer to the full-sample estimator. 

It is easier to understand the issue using a simple example. Consider estimating  the mean $\mathbb EY_t$ using iid data $Y_t$. The full-sample estimator $\widehat\mu$ is the solution to
$$
\widehat\mu=\arg\min_{\mu}L(\mu),\quad \text{where } L(\mu)=\sum_{s=1}^T(Y_s- \mu)^2.
$$
Now consider the ALOO version of this problem. Our definition of $L^{(-t)}(\mu)$ is \textit{not} dropping $Y_t$, but replacing $(Y_t-\mu)^2$ with its expectation:
$$
\breve{\mu}^{(-t)}=\arg\min_{\mu}L^{(-t)}(\mu),\quad \text{where } L^{(-t)}(\mu)=\sum_{s\neq t}(Y_s- \mu)^2+ \mathbb E (Y_t-\mu)^2.
$$
The solution is then $\breve{\mu}^{(-t)}= \frac{1}{T}(\sum_{s\neq t}Y_s+\mathbb EY_t)$. Then straightforward calculations can verify  that $\breve{\mu}^{(-t)}$ (although infeasible) is more efficient and  ``closer" to the full-sample average $\widehat\mu$ than the naive dropping-$t$ estimator $\bar Y_{-t}:=\frac{1}{T-1}\sum_{s\neq t} Y_s$.  For instance,
$$
\frac{\Var (\breve{\mu}^{(-t)} )}{\Var(\bar Y_{-t})} =\left(\frac{T-1}{T}\right)^2<1,\quad \frac{\mathbb E(\breve{\mu}^{(-t)}- \widehat\mu)^2}{\mathbb E(\bar Y_{-t}-\widehat\mu )^2} =\frac{T-1}{T}<1.
$$

The definitions of $L^{(-t)}(B, F)$ and $L^{(-t)}(\mu)$  also 
fulfill  the intuition of the EM algorithm, which imputes the missing data in the loss function by its conditional expectations    before optimizations, rather than simply dropping the missing values.

\subsubsection{Singular vector estimation is unbiased}\label{subsubsec:absorbeffect}

From Algorithm \ref{alg:estimation}, we see that there is no explicit debias step. In fact, in terms of estimating the singular vector space, the singular vector estimator from the least square estimation following the nuclear norm penalization, $\widehat{F}_t$, is unbiased (up to a rotation).

To see this, note that the estimation of $F_t$ has the following maximization problem:
\[\widehat{F}_t \coloneqq \argmax_{f \in \mathbb{R}^K} Q_t(f, \widetilde{\beta})\]
where $Q_t(f,B) = -N^{-1}\sum_{j=1}^N \omega_{jt}(y_{jt}-f' b_j)^2$, $B=(b_1,\dots,b_N)^{\prime}$ and $b_j$ are $K$ dimensional vectors. In this step, $\beta$ is the nuisance parameter and $F_t$ is the parameter of interest. By Taylor expansion, we have, for some invertible matrix $A$,
\begin{align}\label{eq:taylor}
		&\sqrt{N} (\widehat{F}_t-H_1^{-1}F_t) \nonumber\\
		&\quad = - \sqrt{N} A^{-1} \frac{\partial Q_t(H_1^{-1}F_t, \beta H_1)}{\partial f}-\underbrace{\sqrt{N} A^{-1}\frac{\partial^2 Q_t (H_1^{-1}F_t, \beta H_1)}{\partial f \partial \vect(B)}\vect(\widetilde{\beta}-\beta H_1)}_{d} + o_P(1).
	\end{align}
The first term is the score which leads to the asymptotic normality and the second term represents the effect of the $\beta$ estimation which is subject to the shrinkage bias. The second term, while is the ``usual bias" in  a generic machine learning inference problem,  can be 
shown to take the form:
$$
d=\sqrt{N} \varphi H_1^{-1} F_t + o_P(1)
$$
for some $\varphi=o_P(1)$. It has a useful feature of being on the space of $F_t$. Making use of this fact, \eqref{eq:taylor} can be re-written as follows:
\[
\sqrt{N} (\widehat{F}_t-H_2 F_t) = - \underbrace{\sqrt{N} A^{-1} \frac{\partial Q_t(H_1^{-1}F_t, \beta H_1)}{\partial f}}_{\text{asymptotically normal}} + o_P(1)
	\]
by defining $H_2 \coloneqq (I_K + \varphi) H_1^{-1}$. Note that the non-negligible bias term in $d$ is absorbed by the rotation matrix $H_2$, and thus $\widehat{F}_t$ can unbiasedly estimate $F_t$ up to this new rotation. Then, in Step 4 of Algorithm \ref{alg:estimation}, $\widehat{\beta}$, the least square estimator using $\widehat{F}$ as a regressor, can unbiasedly estimate $\beta_i$ up to the rotation since $\widehat{F}_t$ has only a higher order bias now. As a result, the product of them estimates $M_{it}$ unbiasedly:
\begin{align*}          
\widehat{M}_{it}= \widehat{\beta}_{i}' \widehat{F}_t &\approx \beta_{i}' H_2^{ -1}H_2 F_t  = M_{it}
\end{align*}
which allows us to conduct inference successfully.
This is how the two-step least squares procedure works.  

	\section{Asymptotic Results} \label{sec:result}
	
\subsection{Inferential theory}
	
	This section presents the inferential theory. We provide the asymptotic normality of the estimator of the group average of $M_{it}$. Our assumptions allow  the   rank $K$  to grow, but  slowly.  Remind the following notation:
	\[
	h_t( \zeta_{i}) = \sum_{r=1}^{K} \kappa_{t,r}  \phi_{r}( \zeta_{i}) + M^{R}_{it} = \beta_{i}^{\prime}F_t   + M^{R}_{it}, 
	\]
	where $\beta_{i} = (\phi_{1}( \zeta_{i}),\dots, \phi_{K}( \zeta_{i}))^{\prime}$ and $F_t = (\kappa_{t,1},\dots,\kappa_{t,K})^{\prime}$. Let $S_{\beta} = N^{-1}\sum_{i=1}^N \beta_i \beta_i'$, $S_{F} = T^{-1}\sum_{s=1}^T F_s F_s'$, and $Q = S_{\beta}^{1/2}S_{F}^{1/2}$.
	
	\begin{assumption}[Sieve representation]\label{asp:nonparametric_sieve}
		\textit{(i) $\{h_{t}(\cdot) \}_{t\leq T}$ belong to ball $\calH\left( \calZ,\norm{\cdot }_{L_2},C  \right)$ inside a Hilbert space spanned by the basis $\{\phi_r \}_{r\geq 1}$, with a uniform $L_2$-bound $C$: 
			$ \sup_{h\in \calH( \calZ,\norm{\cdot}_{L_2})}\|h\| \leq C,$ where $\calZ$ is the support of $\zeta_{i}$.\\
			(ii) The sieve approximation error satisfies: For some $\nu>0$, $\max_{i,t}|M^{R}_{it}| \leq C K^{-\nu}$.\\
			(iii) For some $C>0$, $\max_{r \leq K} \sup_{\zeta} |\phi_r (\zeta)| < C$. In addition, there is $\eta > 0$ such that $\psi_{\min}^{-1}\left( S_{\beta} \right) < \eta$ and $\psi_{\min}^{-1}\left( S_F \right) < \eta$ with probability converging to 1. \\
			(iv) $ (NT)^{-1}\sum_{i,t} h_t^2 ( \zeta_{i}) = O_P (1)$.\\
			(v) There are constants $\delta,g \geq 0$ such that
			$\psi_{1}(Q) / \psi_{K}(Q) = O_P(K^{\delta})$, $\min_{1\leq r \leq K-1} \psi_{r}(Q) - \psi_{r+1}(Q) \geq c K^{-g}$ for some constant $c>0$.
			}
	\end{assumption}
	
	First, we present some assumptions for the sieve representation. Assumption \ref{asp:nonparametric_sieve} (ii) is well satisfied with a large $\nu$ if the functions $\{h_t\left( \cdot \right)\} $ are sufficiently smooth. For example, consider $h_t$ belonging to a H$\ddot{o}$lder class: for some $a,b,C>0$, $\left\lbrace h : \|D^{b}h(x_1) - D^{b}h(x_2) \| \leq C \| x_1 - x_2\|^{a}\right\rbrace.$ In addition, suppose that we take a usual basis like polynomials, trigonometric polynomials, and B-splines. Then, $\max_{i,t}|M^{R}_{it}| \leq C K^{-\nu},$ and $\nu = 2(a + b)/\text{dim}(\zeta_i).$ So, Assumption \ref{asp:nonparametric_sieve} (ii) is satisfied with very large $\nu$ if $\{h_t\left( \cdot \right)\} $ are smooth. In addition, the first part of Assumption \ref{asp:nonparametric_sieve} (iii) can be satisfied if the basis is a bounded basis like trigonometric basis or $\zeta_i$ has a compact support. Assumption \ref{asp:nonparametric_sieve} (iv) and (v) are not  restrictive, and have been verified by \cite{chernozhukov2021inference}.

\begin{assumption}[DGP for $\varepsilon_{it}$ and $\omega_{it}$]\label{asp:nonparametric_dgp}
	\textit{(i) Conditioning on $\calM$, $\varepsilon_{it}$ is zero-mean, sub-gaussian random variable such that $\bbE[\varepsilon_{it} | \mathcal{M}] = 0$, $\bbE[\varepsilon_{it}^2|\mathcal{M}] = \sigma_{it}^2 \leq \sigma^2$, $ \bbE [\exp(s \varepsilon_{it})|\mathcal{M}] \leq \exp(C s^2 \sigma^2)$, $\forall s \in \bbR$ for some constant $C>0$. We assume that $\sigma^2$ is bounded above and $\sigma^2_{it}$ are bounded away from zero. In addition, $\varepsilon_{it}$ is indepedent across $i$ and $t$.\\
		(ii) $\Omega$ is independent of $\calE$. Conditioning on $\calM$, $\omega_{it}$ is independent across $t$. In addition, $\bbE[\omega_{it}|\calM]=\bbE[\omega_{it}] = p_i$ where $0 < p_{\min} \leq p_i \leq p_{\max} \leq 1$.\\
		(iii) Let $a_t$ be the column of either $\Omega- \Pi \bold{1}_N\bold{1}^{\prime}_T$ or $\Omega \circ \calE.$ Then, $\{a_t\}_{t \leq T}$ are independent sub-gaussian random vector with $\bbE[a_t]=0$; more specifically, there is $C>0$ such that
		\[         
			\max_{t\leq T} \sup_{\|x\|=1} \bbE[\exp(sa_t^{\prime}x)] \leq  \exp(s^2C), \quad \forall s \in \bbR.
		\]		}
\end{assumption}

We assume the heterogeneous observation probability across $i$. It generalizes the homogeneous observation probability assumption which is a typical assumption in the matrix completion literature. The sub-gaussian assumption in Assumption \ref{asp:nonparametric_dgp} (iii) helps us to bound $\norm{\Omega \circ \calE }$ and $\norm{\Omega- \Pi \bold{1}_N\bold{1}_T^{\prime}}$.

While the serial independence of the missing data indicators $\omega_{it}$ is assumed, we allow they are cross-sectional dependence among $i$. In doing so, we assume a cluster structure in $\{1, \dots, N\}$, i.e., there is a family of nonempty disjoint clusters, $\calC_1,\dots, \calC_{\rho}$ such that $\cup_{g=1}^{\rho} \calC_g = \{1, \dots, N\}$. So we divide units $\{1,...,N\}$ into $\rho$ disjoint clusters. In addition, denote the size of the largest cluster by $\vartheta$. That is, $\vartheta=\max_{g} |\calC_g|_o$. We highlight that $\vartheta$ is allowed to increase as $N$ and $T$ increase. 
 
\begin{assumption}[Cross-sectional Dependence in $\omega_{it}$]\label{asp:clusterdependence}
	\textit{Cross sectional units $\omega_{it}$ are independent across clusters. Within the same cluster, arbitrary dependence is allowed, but overall, we require \\$\max_t \max_i \sum_{j=1}^N |\Cov(\omega_{it},\omega_{jt}|\calM) | < C.$}
\end{assumption}

Due to the cluster structure in Assumption \ref{asp:clusterdependence} (i), we can construct a ``leave-cluster-out'' estimator $\breve{\beta}^{\{-i\}}$ which is independent of the sample of unit $i$. Similarly to the idea of \eqref{eq:nonconvexloss} and \eqref{eq:leaveoneoutloss}, we can rule out the samples of the cluster that includes unit $i$. The difference from \eqref{eq:leaveoneoutloss} is that we identify all the units which are in the same cluster as unit $i$ and replace their rows of the goodness of fit part by their conditional expectations.\footnote{For the formal definitions of the estimators, please refer to Section \ref{sec:estimator_definition} of Appendix and Remark \ref{rem:simplenotation} in the section.} Together with the leave-one-out estimator $\breve{\beta}^{(-t)}$, the leave-cluster-out estimator $\breve{\beta}^{\{-i\}}$ plays a pivotal role in showing the solution of \eqref{eq:convexob} is close to that of \eqref{eq:nonconvexloss}.

The parameter for the cluster size $\vartheta$ is bounded by Assumption \ref{asp:nonparametric_parameters}. For instance, in the case where $N \asymp T$ and $\{h_{t}(\cdot) \}_{t\leq T}$ are smooth enough, if we estimate the cross-sectional average of a certain period, the assumption requires $\vartheta \approx o(\sqrt{N / \log N} )$ since $K$ is allowed to grow very slowly when $\{h_{t}(\cdot) \}_{t\leq T}$ are smooth.

We are interested in making inference about group-averaged effects. Let $\mathcal G $ be a particular group; the object of interest is 
$$
\frac{1}{|\calG|_o}\sum_{(i,t)\in\calG}  {M}_{it}=\frac{1}{|\calG|_o}\sum_{(i,t)\in\calG}  h_t(\zeta_i).
$$
Here the group of interest as $\calG = \calI \times \calT$ where $\calI \subseteq \{1, \ldots, N\}$ and $\calT \subseteq \{1, \ldots, T\}$. We impose the following assumption on the rates of parameters. Define a sequence $\psi_{NT}$ as $\psi_{NT} \asymp \sqrt{K^{-(2\delta +1)} \sum_{i=1}^N \sum_{t=1}^Th_t^2(\zeta_i)}$. It is a lower bound of $\psi_{\min}(\beta F')$ and works as the parameter for signal. Recall that $K$ denotes the sieve dimension.
	\begin{assumption}[Parameter size and signal-to-noise ratio]\label{asp:nonparametric_parameters}
		\textit{Let $\gamma = \frac{p_{\max}}{p_{\min}}$ and $\tilde{\vartheta} = \max\{\vartheta, \log N + \log T\}$. Then, we have
			\begin{align*}     
				&(i)\ \ \min\{|\calI|_o^{\frac{1}{2}},|\calT|_o^{\frac{1}{2}}\}\ \tilde{\theta}\eta^3 \gamma^{4}K^{(4+2g+\frac{13}{2}\delta)}\max\{\sqrt{N\log N},\sqrt{ T\log T} \} = o(p_{\min}^{\frac{3}{2}} \min\{N,T\}),\quad \quad \quad \quad \quad \quad \quad  \quad \quad \\
				&\qquad \min\{|\calI|_o^{\frac{1}{2}},|\calT|_o^{\frac{1}{2}}\} \eta^{\frac{1}{2}}\gamma^{3}K^{(1+g+\frac{7}{2}\delta)} \max\{N^{\frac{3}{2}},T^{\frac{3}{2}}\} = o(p_{\min}^{\frac{3}{2}} \psi_{NT}^2),\quad \quad \quad \quad \quad  \quad \quad \\
				&(ii)\ \ 
				\min\{|\calI|_o^{\frac{1}{2}},|\calT|_o^{\frac{1}{2}}\}\eta^{\frac{3}{2}} \gamma^{2} \max\{\sqrt{N},\sqrt{T} \}  =   o(p_{\min}^{\frac{1}{2}}K^{(\nu-2\delta-\frac{3}{2})}),\\
				&\qquad
				\min\{|\calI|_o^{\frac{1}{2}},|\calT|_o^{\frac{1}{2}}\} \eta^{\frac{1}{2}} \gamma^{\frac{3}{2}} \max\{\sqrt{N},\sqrt{T} \} \sqrt{NT} =   o(\psi_{NT} p_{\min}^{\frac{1}{2}} K^{(\nu-\delta-\frac{1}{2})}).
			\end{align*}}
	\end{assumption}
	
Assumption \ref{asp:nonparametric_parameters} (ii) is used to bound the sieve approximation error. For this condition to be satisfied, the smoothness of $\{h_{t}(\cdot) \}_{t\leq T}$ is crucial. If $\{h_{t}(\cdot) \}_{t\leq T}$ are smooth enough, $\nu = 2(a + b)/\text{dim}(\zeta_i)$ can be arbitrarily large. Hence, Assumption \ref{asp:nonparametric_parameters} (ii) can be easily satisfied with a slowly increasing $K$ as long as $\{h_{t}(\cdot) \}_{t\leq T}$ is smooth.
 
Assumptions \ref{asp:nonparametric_parameters} (i) is the conditions about sample complexity and signal-to-noise ratio. As long as $K,\eta,\gamma$ are bounded or increase sufficiently slowly, it would be satisfied. Note that, in the cases like the cross-sectional average of a certain period t or the time average of a certain unit i, $ \min\{|\calI|_o^{\frac{1}{2}},|\calT|_o^{\frac{1}{2}}\} =1$. In many interesting cases, $\min\{|\calI|_o^{\frac{1}{2}},|\calT|_o^{\frac{1}{2}}\}$ is usually not that large. However, due to Assumption \ref{asp:nonparametric_parameters} (i), we cannot derive the inferential theory in the case where both $|\calI|_o$ and $|\calT|_o$ are large like $|\calI|_o=N$ and $|\calT|_o=T$. In this case, the asymptotically normal distribution part cannot dominate other residual parts, since the convergence rate of the asymptotically normal distribution part is roughly $\frac{1}{\sqrt{N|\calT|_o}} + \frac{1}{\sqrt{T|\calI|_o}}$, while that of the residual term is similar to or greater than $\frac{1}{\sqrt{NT}}$ regardless of the group size. For inference, at least one part of the asymptotically normal term should dominate other residual terms. On the other hand, in terms of the convergence rate, the large sizes of $|\calI|_o$ and $|\calT|_o$ are beneficial, as noted in Section \ref{sec:convergencerate} in Appendix. In addition, for comparison with the conditions of other low-rank literature, it would be helpful to refer to Assumption \ref{asp:smallop1} in Appendix where we consider the general low-rank model.

Under the above assumptions, Theorem \ref{thm:generalfactor_groupclt} shows that the estimator for the group average of $M_{it}$ has the asymptotic normality:
$$\calV_{\calG}^{-\frac{1}{2}}\left( \frac{1}{|\calG|_o}\sum_{(i,t)\in\calG}  \widehat{M}_{it} - \frac{1}{|\calG|_o}\sum_{(i,t)\in\calG}  M_{it} \right) \conD \calN(0,1),$$
where the asymptotic variance $\calV_{\calG}$ is given in the statement of Theorem \ref{thm:generalfactor_groupclt}, and needs to be estimated.  In this result, $\calG$ can consist of either   multiple columns with multiple rows or solely a certain $(i,t)$, implying that we can conduct inference for one specific element of the matrix.  
	
	  To make the estimation of $\calV_{\mathcal G}$  feasible,   we consider the case of $\bbE[\varepsilon_{it}^2|\calM] = \sigma^2_{i}$. Let $U_{i}'$ is the $i$-th row of the left singular vector of $\beta F'$ and $V_{t}'$ is the $t$-th row of the right singular vector of $\beta F'$. The following theorem gives the feasible asymptotic normality. 
	
	\begin{theorem} [Feasible CLT]\label{thm:feasibleclt}
		Suppose Assumptions \ref{asp:nonparametric_sieve} - \ref{asp:nonparametric_parameters} hold. In addition, suppose that\\ $\norm{\frac{\sqrt{N}}{|\calI|_o}\sum_{i \in \calI}U_{M^*,i}} \geq c$ and $\norm{\frac{\sqrt{T}}{|\calT|_o}\sum_{t \in \calT}V_{M^*,t}} \geq c$ for some constant $c>0$. Then we have
		\begin{align*}
			\widehat{\calV}_{\calG}^{-\frac{1}{2}}\left( \frac{1}{|\calG|_o}\sum_{(i,t)\in\calG}  \widehat{M}_{it}  - \frac{1}{|\calG|_o}\sum_{(i,t)\in\calG}  M_{it} \right) \conD \calN(0,1), 
		\end{align*}
		where
\begin{align*}
  \widehat{\calV}_{\calG}&=\frac{1}{|\calT|_o^2} \sum_{t\in\calT}\widehat{\bar{\beta}}_{\calI}^{\prime}\left( \sum_{j=1}^{N} \omega_{jt}\widehat{\beta}_{j}\widehat{\beta}_{j}^{\prime} \right)^{-1}\left( \sum_{j=1}^{N} \omega_{jt} \widehat{\sigma}^2_{j} \widehat{\beta}_{j}\widehat{\beta}_{j}^{\prime} \right) \left( \sum_{j=1}^{N} \omega_{jt}\widehat{\beta}_{j}\widehat{\beta}_{j}^{\prime} \right)^{-1}\widehat{\bar{\beta}}_{\calI} \\
  & \ \ + \frac{1}{|\calI|_o^2} \sum_{i\in\calI} \widehat{\sigma}^2_i \widehat{\bar{F}}_\calT^{\prime} \left( \sum_{s=1}^{T}\omega_{is}\widehat{F}_s \widehat{F}_s^{\prime}  \right)^{-1}\widehat{\bar{F}}_\calT ,
\end{align*}
$\widehat{\bar{\beta}}_{\calI} = \frac{1}{|\calI|_o}\sum_{a \in \calI}\widehat{\beta}_{a}$, $\widehat{\bar{F}}_\calT = \frac{1}{|\calT|_o}\sum_{a \in \calT}\widehat{F}_{a}$, $\widehat{\sigma}_i^2 =\frac{1}{|\calW_i|_o}\sum_{t\in\calW_i} \widehat{\varepsilon}_{it}^2$, $\calW_i = \{t:\omega_{it}=1\}$ and $\widehat{\varepsilon}_{it} = y_{it} - \widehat{\beta}_{i}^{\prime}\widehat{F}_t$.
\end{theorem}

\subsection{Semiparametric efficiency}\label{sec:efficiency}
 
We now establish the semiparametric efficiency of our estimator, following a similar approach as in \cite{jankova2018semiparametric}. In order to make calculation tractable, we suppose that $\omega_{it}\sim \mathrm{Bernoulli}(p)$ and $\varepsilon_{it}\sim \calN(0,\sigma^2)$ are independent across $(i,t)$. We will focus on the case of block group, where both $|\calI|_o$ and $|\calT|_o$ are finite or growing slowly, satisfying $N|\calT|_o \ll T^2 |\calI|_o^2$ and $T|\calI|_o \ll N^2 |\calT|_o^2$. The other cases like cross-sectional and serial groups (e.g., $|\calI|_o =N$ and $|\calT|_o$ is finite or slowly growing, or vice versa) can also be attained, which are very similar to Theorem 4.2 in \cite{chernozhukov2021inference}. Hence, we omit them. The novelty of our  efficiency theorem is that  it is  for estimating   the general block group.

As specified in Theorem \ref{thm:generalfactor_groupclt}, the asymptotic variance  in this case is 
\begin{align*}
    \calV_{\calG} &=  \frac{\sigma^2}{|\calT|_o^2} \sum_{t \in \calT} \bar{\beta}_{\calI}' \left( \sum_{j =1}^N \omega_{jt} \beta_j \beta_j'\right)^{-1} \bar{\beta}_{\calI} + \frac{\sigma^2}{|\calI|_o^2} \sum_{i \in \calI} \bar{F}_{\calT}' \left( \sum_{s =1}^T \omega_{is} F_s F_s'\right)^{-1} \bar{F}_{\calT} \cr 
    &=s_*^2(M, p, \sigma) + o(s_*^2(M, p, \sigma)) \cr
   s^2_*(M,p, \sigma)  &:= \frac{\sigma^2}{p} \frac{1}{|\calT|_o}\bar{\beta}_{\calI}'(\beta'\beta)^{-1}\bar{\beta}_{\calI} + \frac{\sigma^2}{p} \frac{1}{|\calI|_o}\bar{F}_{\calT}'(F'F)^{-1}\bar{F}_{\calT}.
\end{align*}

 The following theorem shows that $s^2_*(M,p, \sigma)$ is the asymptotic Cram\'er-Rao bound for asymptotically unbiased estimators. 
 
\begin{theorem}\label{thm:efficiency}
    Suppose $\omega_{it} \sim \mathrm{Bernoulli}(p)$ and $\varepsilon_{it}\sim \calN(0,\sigma^2)$ are independent across $(i,t)$. Suppose also that $N|\calT|_o \ll T^2 |\calI|_o^2$ and $T|\calI|_o \ll N^2 |\calT|_o^2$. Define $$\calA = \{(M,p, \sigma) : M=M^{\star}+M^R, M^{\star}=\beta F', \mathrm{rank}(M^{\star})\leq K, \,\, \text{and Assumptions \ref{asp:nonparametric_sieve}-\ref{asp:nonparametric_parameters} hold}\}.$$
    Let $U (Y, \Omega)$ be an asymptotically unbiased estimator of $|\calG|^{-1} \sum_{(i,t) \in \calG}M_{it}$ in that $$\bbE_{M,  p, \sigma}U (Y, \Omega) - |\calG|^{-1} \sum_{(i,t) \in \calG}M_{it} = o(s_*(M,p,\sigma))$$ where $\bbE_{M,p, \sigma}$ denotes the expectation with respect to given $(M,  p, \sigma)$. Then for any sequence of $(M,p, \sigma) \in \calA$, we have
     \begin{align*}
   \liminf_{N,T \rightarrow \infty}\frac{ \bbE_{M,p, \sigma}\left[U (Y, \Omega)-|\calG|^{-1} \sum_{(i,t) \in \calG}M_{it}\right]^2}{s^2_*(M,p, \sigma)} \geq 1,
\end{align*}  
with probability converging to 1.
\end{theorem}

	\section{Applications to Heterogeneous Treatment Effect Estimation}\label{sec:treatment}
	
	In this section, we propose the inference procedure for treatment effects by utilizing the asymptotic results in Section \ref{sec:result}. Following the causal potential outcome setting (e.g., \cite{rubin:1974}, \cite{imbens:2015}), we assume that for each of $N$ units and $T$ time periods, there exists a pair of potential outcomes, $y_{it}^{(0)}$ and $y_{it}^{(1)}$ where $y_{it}^{(0)}$ denotes the potential outcome of the untreated situation and $y_{it}^{(1)}$ denotes the potential outcome of the treated situation. Importantly, among potential outcomes $y_{it}^{(0)}$ and $y_{it}^{(1)}$, we can observe only one realized outcome $y_{it}^{(\Upsilon_{it})}$ where $\Upsilon_{it} = 1\{\text{unit $i$ is treated at period $t$}\}$. Hence, we have two incomplete potential outcome matrices, $Y^{(0)}$ and $Y^{(1)}$, having missing components, and the problem of estimating the treatment effects can be cast as a matrix completion problem because of the missing components in the two matrices. 
	
	Specifically, we consider the nonparametric model such that for each $\iota \in \{0,1\}$,
	\[y_{it}^{(\iota )} = M_{it}^{(\iota )} + \varepsilon_{it}
	= h_{t} ^{(\iota )} (\zeta_{i}) + \varepsilon_{it}, \]       
	where $\varepsilon_{it}$ is the noise, $\zeta_{i}$ is a vector of unit specific latent state variables. We regard $h_t^{(\iota)}(\cdot)$ as a deterministic function while $\zeta_{i}$ is a random vector. In the model, the treatment effect comes from the difference between the time-varying treatment function $h_t^{(1)}(\cdot)$ and the control function $h_t^{(0)}(\cdot)$. Let $\omega_{it}^{(\iota )} =1{\{y_{it}^{(\iota )}  \text{ is observed}\}}$. Then, $\omega_{it}^{(1)} = \Upsilon_{it}$ and $\omega_{it}^{(0)} = 1 - \Upsilon_{it}$ because we observe $y_{it}^{(1)}$ when there is a treatment on $(i,t)$ and observe $y_{it}^{(0)}$ when there is no treatment on $(i,t)$.
	
    We suppose the following seive representation for $h_t^{(\iota)}$ :
	\[ h_t^{(\iota)} ( \zeta_{i})= \sum_{r=1}^{K} \kappa_{t,r}^{(\iota)}  \phi_{r}( \zeta_{i}) + M^{R(\iota)}_{it},\ \ \ \ \ \ \iota \in \{0,1\} \]
	where $\kappa^{(\iota)}_{t,r}$ is the sieve coefficient, $\phi_{r}( \zeta_{i})$ is the sieve transformation of $\zeta_i$ using the basis function $\phi_{r}( \cdot)$ and $M_{it}^{R(\iota)}$ is the sieve approximation error. Then, by representing $\sum_{r=1}^{K} \kappa_{t,r}^{(\iota)}  \phi_{r}( \zeta_{i}) $ as $\beta_{i}^{\prime}F_t^{(\iota)}$ where $\beta_{i} = [\phi_{1}( \zeta_{i}),\dots, \phi_{K}( \zeta_{i})]^{\prime}$ and $F_t^{(\iota)} = [ \kappa^{(\iota)}_{t,1},\dots,\kappa^{(\iota)}_{t,K}]^{\prime}$, $h^{(\iota)}_t(\zeta_{i})$ can be successfully represented as the approximate factor structure.
	
We make inference about the average treatment effect for a particular group of interest $(i,t)\in\calG$: 
$$\frac{1}{|\calG|_o}\sum_{(i,t)\in\calG} \Gamma_{it},\quad\text{where } \Gamma_{it} = M_{it}^{(1)} - M_{it}^{(0)}.$$
The individual treatment effect $\Gamma_{it}$ is estimated by   $\widehat{\Gamma}_{it} = \widehat{M}_{it}^{(1)} -\widehat{M}_{it}^{(0)}$ where $\widehat{M}_{it}^{(0)}$ and $\widehat{M}_{it}^{(1)}$ are estimators of $M_{it}^{(0)}$ and $M_{it}^{(1)}$, respectively.  
Hence, by implementing the estimation steps in Algorithm \ref{alg:estimation} for each $\iota \in \{0,1\}$, we can derive the estimators for the group average of $M_{it}^{(0)}$ and $M_{it}^{(1)}$, and construct the average treatment effect estimator.
	
	The notations are essentially the same as those in Section \ref{sec:modelandestimation}, and we just put the superscript $(\iota)$ to all notations to distinguish the pair of potential realizations.

	\begin{theorem}[Feasible CLT] \label{cor:feasibletreatmentclt}
			Suppose the assumptions of Theorem \ref{thm:feasibleclt} hold for each $\iota \in \{0,1\}$. With $\bbE[\varepsilon_{it}^2|\calM] = \sigma^2_{i}$, we have
		\begin{gather*}
			\left( \widehat{\calV}_{\calG}^{(0)} + \widehat{\calV}_{\calG}^{(1)}\right)^{-\frac{1}{2}} \left( \frac{1}{|\calG|_o}\sum_{(i,t)\in\calG}  \widehat{\Gamma}_{it}
			- \frac{1}{|\calG|_o}\sum_{(i,t)\in\calG} \Gamma_{it} \right)  \conD \calN(0,1),
		\end{gather*}
  where for each $\iota \in \{0,1\}$,
\begin{align*}
  \widehat{\calV}_{\calG}&=\frac{1}{|\calT|_o^2} \sum_{t\in\calT}\widehat{\bar{\beta}}_{\calI}^{(\iota)\prime}\left( \sum_{j=1}^{N} \omega_{jt}^{(\iota)} \widehat{\beta}_{j}^{(\iota)}\widehat{\beta}_{j}^{(\iota)\prime} \right)^{-1}\left( \sum_{j=1}^{N} \omega_{jt}^{(\iota)} \widehat{\sigma}^{(\iota)2}_{j} \widehat{\beta}_{j}^{(\iota)} \widehat{\beta}_{j}^{(\iota)\prime} \right) \left( \sum_{j=1}^{N} \omega_{jt}^{(\iota)} \widehat{\beta}_{j}^{(\iota)} \widehat{\beta}_{j}^{(\iota)\prime} \right)^{-1}\widehat{\bar{\beta}}^{(\iota)}_{\calI}\\
  &\ \ +  \frac{1}{|\calI|_o^2} \sum_{i\in\calI} \widehat{\sigma}_i^{(\iota)2}\widehat{\bar{F}}_\calT^{(\iota)\prime} \left( \sum_{s=1}^{T}\omega_{is}^{(\iota)} \widehat{F}_s^{(\iota)}  \widehat{F}_s^{(\iota)\prime}  \right)^{-1}\widehat{\bar{F}}_\calT^{(\iota)} .
\end{align*}
Here, $\widehat{\bar{\beta}}_{\calI}^{(\iota)} = \frac{1}{|\calI|_o}\sum_{a \in \calI}\widehat{\beta}_{a}^{(\iota)} $, $\widehat{\bar{F}}_\calT^{(\iota)}  = \frac{1}{|\calT|_o}\sum_{a \in \calT}\widehat{F}_{a}^{(\iota)} $, $\left( \widehat{\sigma}_i^{(\iota)} \right) ^2 =\frac{1}{|\calW_i^{(\iota)} |_o}\sum_{t \in \calW_i^{(\iota)} } \left( \widehat{\varepsilon}_{it}^{(\iota)} \right) ^2$, $\calW_i^{(\iota)}  = \{t:\omega_{it}^{(\iota)} =1\}$ and $\widehat{\varepsilon}_{it}^{(\iota)}  = y_{it}^{(\iota)}  - \widehat{\beta}_{i}^{(\iota)\prime}\widehat{F}_t^{(\iota)} $.
\end{theorem}

\section{Empirical study: Impact of the president on allocating the U.S. federal budget to the states}\label{sec:empirical}

To illustrate the use of our inferential theory, we present an empirical study about the impact of the president on allocating the U.S. federal budget to the states. The allocation of the federal budget in the U.S. is the outcome of a complicated process involving diverse institutional participants. However, the president plays a particularly important role among the participants. Ex-ante, the president is responsible for composing a proposal, which is to be submitted to Congress and initiates the actual authorization and appropriations processes. Ex post, once the budget has been approved, the president has a veto power that can be overridden only by a qualified majority equal to two-thirds of Congress. In addition, the president exploits extra additional controls over agency administrators who distribute federal funds.

There is a vast theoretical and empirical literature about the impact of the president on allocating the federal budget to the states (e.g., \cite{cox1986electoral}, \cite{anderson1991congressional}, \cite{mccarty2000presidential}, \cite{larcinese2006allocating}, \cite{berry2010president}). In particular, \cite{cox1986electoral} provide a theoretical model which supports the idea that more funds are allocated where the president has larger support because of the ideological relationship between voters and the president, and \cite{larcinese2006allocating} have found that states which supported the incumbent president in past presidential elections tend to receive more funds empirically. We contribute by showing the impact using our inferential theory for the heterogeneous treatment effect with a wider set of data.

Here, the hypothesis we want to test is whether federal funds are disproportionately targeted to states where the incumbent president is supported in the past presidential election. We use data on federal outlays for the 50 U.S. states with the District of Columbia from 1953 to 2018. The data are obtained from websites of the U.S. Census Bureau, NASBO (National Association of State Budget Officers), and SSA (Social Security Administration). 

Following Section \ref{sec:treatment}, we set the treatment indicator as $\Upsilon_{it} = 1$ if the state $i$ supported the president of year $t$ in the presidential election, and $\Upsilon_{it} = 0$ otherwise. If the candidate whom the state $i$ supported in the previous presidential election is the same as the president at year $t$, we consider it as ``treated" and otherwise, we consider it as ``untreated". 
While applying our inferential procedure, we adopt the assumption that the treatment   (whether state $i$ supported the resident in the election) is exogenously assigned, which is probably not practical, but we take our stand on this  assumption in this study, and do not claim a causal interpretation of the treatment effect.

In addition, for the outcome variable $y_{it}$, we use the following ratio: $y_{it} = (\tilde{y}_{it}/\sum_{i} \tilde{y}_{it}) \times 100$ where $\tilde{y}_{it}$ is the per-capita federal grant in state $i$ at year $t$. Note that the outcome variable, $y_{it}$, is a proportion so that $\sum_i y_{it}=100$ for all $t,$ which is to treat each period equally.

\begin{figure}[htb!]
\begin{center}
    \includegraphics[width=\linewidth]{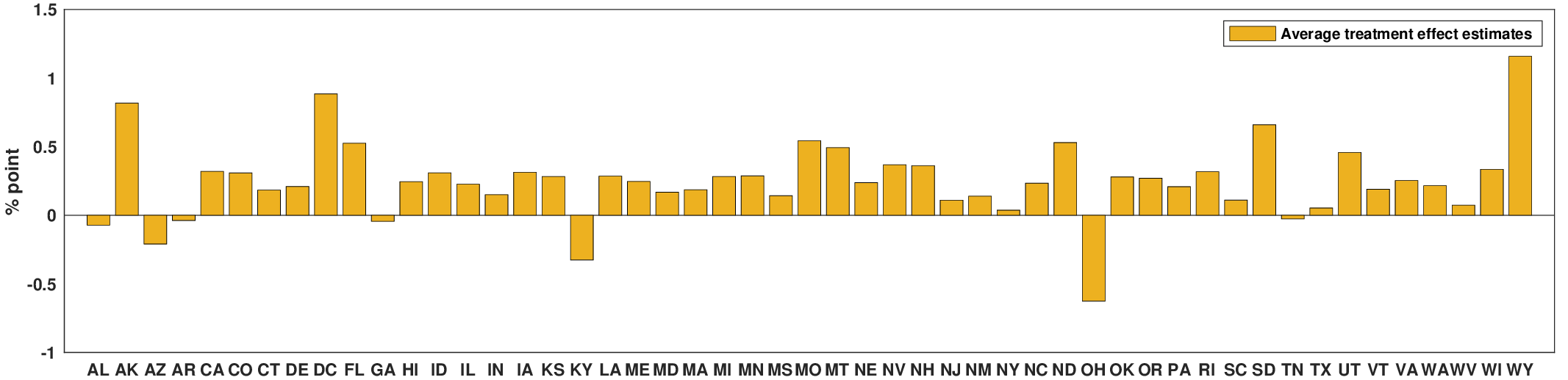}\\
			\includegraphics[width=\linewidth]{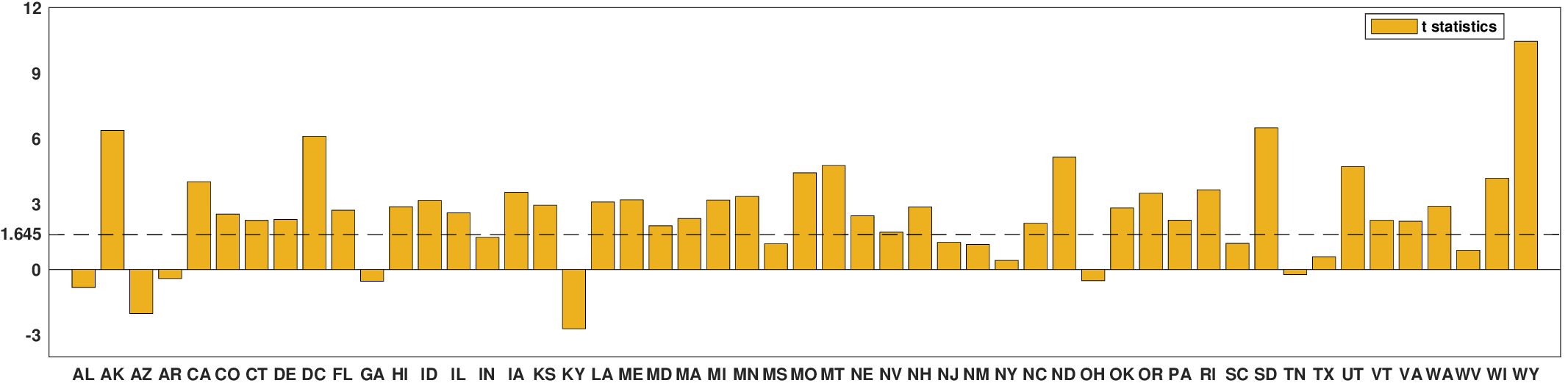}
\end{center}
	\caption{\small  State effects and corresponding t-statistics}	\label{fig:state_main}
	\centering
	\vspace{5mm}
	\begin{minipage}{1\textwidth}
		{\footnotesize NOTE: When we use the \cite{benjamini1995controlling}  procedure to control the size of the false discovery rate at 5\%, the list of states with significant effects is unchanged.}
	\end{minipage}
\end{figure}

Our inferential theory allows novel approaches to study the following effects:
\begin{enumerate}
    \item State Effects: the time average of the treatment effect of each state $i$, i.e., $T^{-1}\sum_{t=1}^T\Gamma_{it}.$
    \item Region Effects: the time average of the treatment effect of each ``Region'', i.e.,
     $$
\frac{1}{|\text{Region}|_0 }\sum_{i\in \text{Region}}\frac{1}{T}\sum_{t=1}^T\Gamma_{it}.$$
    \item Loyal/Swing Effects: the time average of the treatment effect of ``loyal'' and ``swing'' states, e.g.,
     $$
\frac{1}{|\text{Loyal States}|_0 }\sum_{i\in \text{Loyal States}}\frac{1}{T}\sum_{t=1}^T\Gamma_{it}.\quad (\text{see Table \ref{tab:numberofswing} for the definition of ``Loyal States"})
$$
 \vspace{-5mm}
    \item President Effects: the average treatment effect of each president, i.e.,
     $$
\frac{1}{|\mathcal T|_0}\sum_{t\in\mathcal T}\frac{1}{N }\sum_{i=1}^N\Gamma_{it}.\quad (\mathcal T \text{ denotes the period of a given President in Office})
$$
    \item Party Effects: the  average   treatment effect of each Party, i.e.,
    $$
\frac{1}{|\mathcal S|_0}\sum_{t\in\mathcal S}\frac{1}{N }\sum_{i=1}^N\Gamma_{it}.\quad (\mathcal S  \text{ denotes the period of a given Party to which the President belonged})
$$
\end{enumerate}

First, Figure \ref{fig:state_main} presents the State Effects and the corresponding t-statistics. The results suggest significantly positive treatment effects in most states. To investigate the reason of differences, we categorize states according to the number of times a state swung the party it supports in the presidential elections as in Table \ref{tab:numberofswing}. Together with Figure \ref{fig:state_main}, it shows that most states with large t-statistics are in ``Loyal states'' while other states are generally in ``Swing state'' or ``Weak swing state''. It suggests that the treatment effect is closely related to the loyalty of states to parties.

\begin{table}[htb!]
\small
\begin{center}
	\begin{threeparttable}
	\caption{\small  Number of swings of each state}
   \begin{tabular}{lllll}
\toprule
Group             & \# of swing & States \\ \midrule
Loyal states      & 0$\sim$2    & DC, AK, ID, KS, NE, ND, OK, SD, UT, WY \\ \midrule
Weak loyal states & 3$\sim$4    & AZ, CA, CT, IL, ME, MA, MN, NJ, OR, SC, VT, VA, WA, IN, MI, MT, TX \\ \midrule
Weak swing states & 5$\sim$6    & AL, CO, DE, HI, MD, NV, NH, NM, NY, NC, RI, IA, MS, MO, PA, TN, WI \\ \midrule
Swing states      & 7$\sim$     & AR, GA, KY, WV, FL, OH, LA \\ \bottomrule
\end{tabular}
    \label{tab:numberofswing}
	\end{threeparttable}
	\end{center}
\end{table}%

\begin{figure}[htb!]
\begin{center}
    	\includegraphics[width=0.8\linewidth]{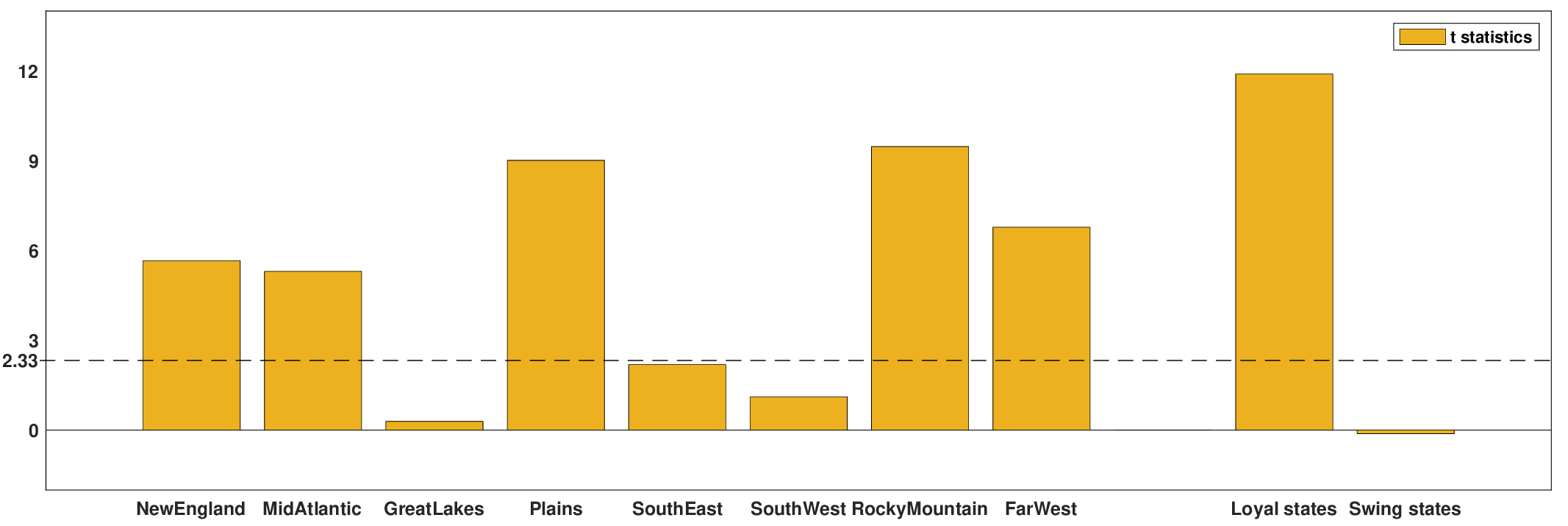}
\end{center}
	\caption{\small  Test statistics for the Region Effects and the Loyal/Swing Effects}	\label{fig:state_region}
\centering
\vspace{5mm}
	\begin{minipage}{1\textwidth}
		{\footnotesize NOTE: ``New England" includes CT, ME, MA, NH, RI,VT, ``Mid Atlantic" includes DE, D.C., MD, NJ, NY, PA, ``Great Lakes" includes IL, IN, MI, OH, WI, ``Plains" includes IA, KS, MN, MO, NE, ND, SD, ``South East" includes AL, AR, FL, GA, KY, LA, MS, NC, SC, TN, VI, WV, ``South West" includes AZ, NM, OK, TX, ``Rocky Mountain" includes CO, ID, MT, UT, WY, and ``Far West" includes AK, CA,HI, NV, OR, WA.}
	\end{minipage}
\end{figure}

In addition, the results for the Region Effects in Figure \ref{fig:state_region} show that, at the 1\% significant level, New England, Mid Atlantic, Plains, Rocky Mountain, and Far West have the positive treatment effects while Great Lakes, South East, and South West do not. Note that Many states in Great Lakes, South East, and South West are in ``Swing states'' or ``Weak swing states.'' As we can see in Figure \ref{fig:state_region}, ``Swing states'' do not have statistically significant positive treatment effects while ``Loyal states'' have significant positive treatment effects. This result is in line with the empirical study of \cite{larcinese2006allocating} finding that states with loyal supports tend to receive more funds, while swing states are not rewarded. In addition, it is aligned with the assertion of \cite{cox1986electoral} that the targeting of loyal voters can be seen as a safer investment as compared to aiming for swing voters and risk-averse political actors may allocate more funds to loyal states.

\begin{figure}[htb!]
\begin{center}
    \includegraphics[width=0.8\linewidth]{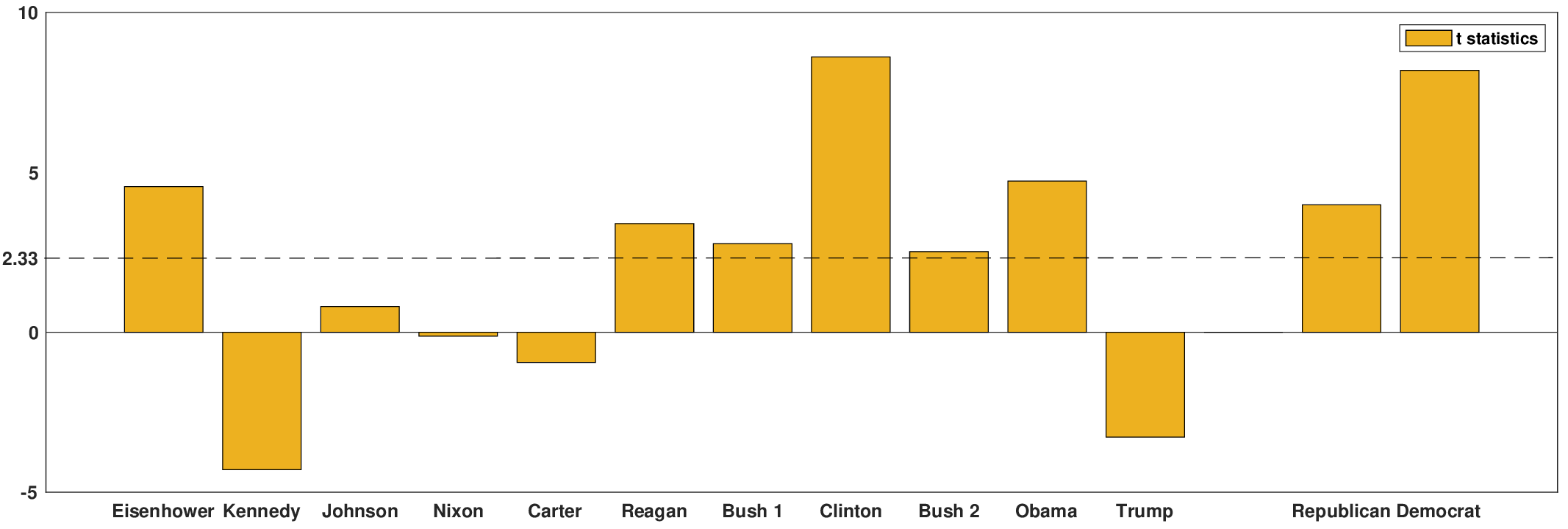}
\end{center}
	\caption{\small     Test statistics for the President Effects and the Party Effects} \label{fig:presidents}
\end{figure}

\begin{figure}[htb!]
\begin{center}
    \includegraphics[width=0.8\linewidth]{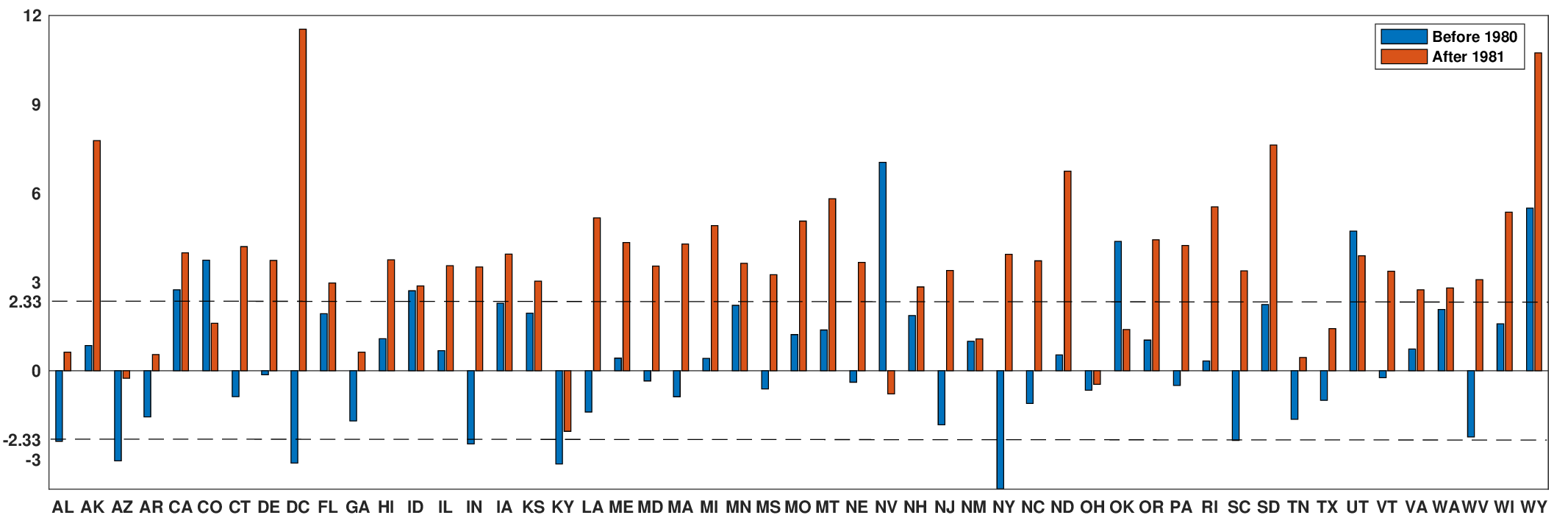}
\end{center}
	\caption{\small  Test statistics for the average treatment effect before 1980 and after 1981}
	\label{fig:beforeafter1980}
\end{figure}

Figure \ref{fig:presidents} shows the President Effects and the Party Effects. Despite some exceptions, there are no statistically significant positive treatment effects before Carter, while there are significant positive treatment effects after Reagan. Figure \ref{fig:beforeafter1980} shows that before 1980, there is no significant positive treatment effect in most states, while there are significant positive treatment effects in most states after 1981. Hence,   there is a substantial difference between `before 1980' and `after 1981' and the tendency that incumbent presidents reward states that showed their support in the presidential elections became significant after Reagan, that is, after the 1980s. It suggests that after the 1980s, the presidents show more influence on the allocation of federal funds to reward their supporters. Evidence is that starting from the 1980s, all presidents have put forward proposals for the introduction of presidential line-item veto and tried to increase the power of the president to control federal spending.  

Finally, when testing for the treatment effects of   multiple states, the tests  may subject to the issue of multiple testing problems, with undesirable     false discovery rates (FDR). We also address this issue by adopting the procedure  of  \cite{benjamini1995controlling}  to control the FDR at 5\%.  We find that the list of states with  significant  treatment effects is unchanged.

\section{Simulation Study} \label{sec:sim}
	
	This section provides the finite sample performances of the estimators. We first study the performances of the estimators of $M_{it}$ and $|\calG|_o^{-1}\sum_{(i,t)\in\calG}  M_{it}$, and then study performances of the average treatment effect estimators. To save space, some results are relegated to Appendix.
	
	First of all, in order to check the estimation quality of our estimator, we compare the Frobenius norms of the estimation errors for several existing estimators of $M$. Our two-step least squares is labelled as ``TLS". We also consider the debiased nuclear norm penalized estimators from \cite{xia2021statistical}, ``(Hetero) XY,'' and \cite{chen:2019inference}, ``(Hetero) CFMY.'' ``(Hetero)'' represents that they are modified to allow the heterogeneous observation probabilities. The comparison also includes the inverse probability weight based estimator, ``IPW,'' from \cite{xiong2020large}, and the EM algorithm based estimator, ``EM,'' from \cite{jin2021factor}. The plain nuclear norm penalized estimator, ``Plain Nuclear,'' and the TLS estimator using sample splitting, ``TLS with SS,'' are also considered. For the data-generating designs, we consider the following three models:
	\begin{align}\label{eq:dgp}
		\nonumber&\bullet \text{Factor model:}\ \ y_{it} = \beta_{1,i}F_{1,t} +\beta_{2,i}F_{2,t}+ \varepsilon_{it},\quad \text{ where }\ \ \beta_{1,i},F_{1,t},\beta_{2,i},F_{2,t} \sim \calN\left( \frac{1}{\sqrt{2}},1\right), \\
		\nonumber&\bullet \text{Nonparametric model 1:}\ \ y_{it} = h_t\left( \zeta_{i}  \right) + \varepsilon_{it}, \quad \text{ where }\ \ h_t(\zeta) =h_t^{\mathrm{poly}}(\zeta) \coloneqq \sum_{r=1}^{\infty} \frac{|U_{t,r}|}{r^3}\cdot \zeta^r,\\
		&\bullet \text{Nonparametric model 2:}\ \ y_{it} = h_t\left( \zeta_{i}  \right) + \varepsilon_{it}, \quad \text{ where }\ \ h_t(\zeta) =h_t^{\mathrm{sine}}(\zeta) \coloneqq \sum_{r=1}^{\infty} \frac{|U_{t,r}|}{r^3}\sin(r\zeta).
	\end{align}
	Here, $U_{t,r}$ is generated from $\calN(2,1)$ and $\zeta_{i}$ is generated from $\text{Uniform}[0,1]$. In addition, $\varepsilon_{it}$ is generated from the standard normal distribution independently across $i$ and $t$. The observation pattern follows a heterogeneous missing-at-random mechanism where $\omega_{it}\sim \text{Bernoulli}(p_i)$ and $p_i$ is generated from Uniform $[0.3,0.7]$.
 
\begin{table}[htb!]
\small
\begin{center}
	\begin{threeparttable}
	\caption{\small $\|\widehat{M} - M \|_F/\sqrt{NT}$ }
    \begin{tabular}{rrrrrrrrrr}
    \toprule
    Sample size & \multicolumn{3}{c}{N = 100, T =100} & \multicolumn{3}{c}{N = 200, T = 100} & \multicolumn{3}{c}{N = 100, T =200} \\ [2pt]
    Model & Factor & Sine  & Poly  & Factor & Sine  & Poly  & Factor & Sine  & Poly \\[2pt]
    \midrule
TLS & 0.3035 & 0.2129 & 0.2057 & 0.2613 & 0.1871 & 0.1777 & 0.2522 & 0.1831 & 0.1831 \\[2pt]
TLS with SS       & 0.3130  & 0.2152 & 0.2080  & 0.2699 & 0.1893 & 0.1805 & 0.2551 & 0.1835 & 0.1836 \\[2pt]
Plain Nuclear      & 0.5637 & 0.3869 & 0.3745 & 0.4827 & 0.3342 & 0.3334 & 0.4814 & 0.3418 & 0.3433 \\[2pt]
(Hetero) CFMY     & 0.3312 & 0.2230  & 0.2128 & 0.2798 & 0.1916 & 0.183  & 0.2740  & 0.1914 & 0.1917 \\[2pt]
(Hetero) XY       & 0.3870  & 0.2369 & 0.2275 & 0.3185 & 0.1984 & 0.1931 & 0.3104 & 0.2019 & 0.2033 \\[2pt]
IPW      & 0.5280  & 0.2446 & 0.2435 & 0.4994 & 0.2184 & 0.2117 & 0.4254 & 0.1997 & 0.2068 \\[2pt]
EM       & 0.3033 & 0.2134 & 0.206  & 0.2611 & 0.1872 & 0.1777 & 0.2517 & 0.1834 & 0.1832 \\
\bottomrule
     \end{tabular}%
			\label{tab:estimationquality}%
		\vspace{3mm}
			\begin{tablenotes}[flushleft]
				\item {\footnotesize NOTE: `` Sine'' and `` Poly'' refer to the functions $h_t^{\mathrm{sine}}(\zeta)$ and $h_t^{\mathrm{poly}}(\zeta)$, respectively.}
			\end{tablenotes}
		\end{threeparttable}
\end{center}
	\end{table}%
	
Table \ref{tab:estimationquality} reports $\|\widehat{M} - M \|_F/\sqrt{NT}$ averaged over 100 replications. We highlight that the TLS shows the best performance in almost all scenarios. Only the EM is comparable to ours, but it computes much slower since it requires multi-step iterations. In contrast, our proposed method does not iterate. 
Also, our method always outperforms the TLS with SS. The (Hetero) XY and (Hetero) CFMY are slightly worse than ours in this experiment. Lastly, both the IPW and the Plain Nuclear show the worst performances uniformly. The IPW, being non-statistically efficient, is only slightly better than the Plain Nuclear.

Additionally, to show the relative advantage of TLS over TLS with sample splitting, Table \ref{tab:comparisonwithss} reports $(\widehat{M}_{it} - M_{it})^2$ in the case where $T$ is small. Here, we choose $(i,t)$ randomly and fix it during replications.
As we can check in the table, when $T$ is relatively small, the performance of TLS with sample splitting is much worse than that of TLS without sample splitting. Especially, in the factor model, the difference in performance is quite large.

\begin{table}[htb!]
  \small
\begin{center}
	\begin{threeparttable}
	\caption{\small  $(\widehat{M}_{it} - M_{it})^2$ Comparison between TLS and TLS with SS}
    \begin{tabular}{cccccccccc}
    \toprule
    Model & \multicolumn{3}{c}{Factor} & \multicolumn{3}{c}{Sine} & \multicolumn{3}{c}{Poly} \\[2pt]
    Sample Size & TLS  & TLS w/ SS &  Ratio &   TLS  & TLS w/ SS &  Ratio &   TLS  & TLS w/ SS &  Ratio \\[2pt]
    \midrule
    N=100,T=20 & 0.4665 & 2.8951 & 16.1\% & 0.1401 & 0.1702 & 82.3\% & 0.1272 & 0.1894 & 67.2\% \\[2pt]
    N=100,T=40 & 0.2162 & 0.2685 & 80.5\% & 0.0736 & 0.0819 & 89.9\% & 0.0807 & 0.0865 & 93.3\% \\[2pt]
    N=100,T=60 & 0.1111 & 0.1300 & 85.5\% & 0.0603 & 0.0637 & 94.7\% & 0.0538 & 0.0567 & 94.9\% \\
    \bottomrule
    \end{tabular}%
  \label{tab:comparisonwithss}%
\vspace{3mm}
			\begin{tablenotes}[flushleft]
				\item {\footnotesize NOTE: The values are the averaged $(\widehat{M}_{it} - M_{it})^2$ over 1,000 replications. `` Ratio'' denotes the ratio between performances of TLS and TLS with SS. Here, we assume $\omega_{it}\sim \text{Bernoulli}(0.5)$. When $T=20$, the working sample size for the sample splitting is only $10$, which leads to singularity issues in the inverse covariance matrix estimation. As a result, the estimator performs badly in this case.}
			\end{tablenotes}
		\end{threeparttable}
\end{center}
	\end{table}%
 
Second, we study the finite sample distributions for standardized estimates defined as
$(\widehat{M}_{it} - M_{it})/se(   \widehat{M}_{it})$. For comparison, we report the results of the Plain Nuclear and the TLS with SS, in addition to the TLS. For the Plain Nuclear, we use the sample standard deviation obtained from the simulations for $se(   \widetilde{M}_{it} )$ because the theoretical variance of it is unknown. For the TLS with SS, we construct the standard error following \cite{chernozhukov2019inference}. Here, we consider the nonparametric models in \eqref{eq:dgp}. Hereinafter, the number of replications is 1,000, and the sample size is $N=T=200.$
	
	\begin{figure}[htb!]
	\begin{center}
	    \includegraphics[width=\linewidth]{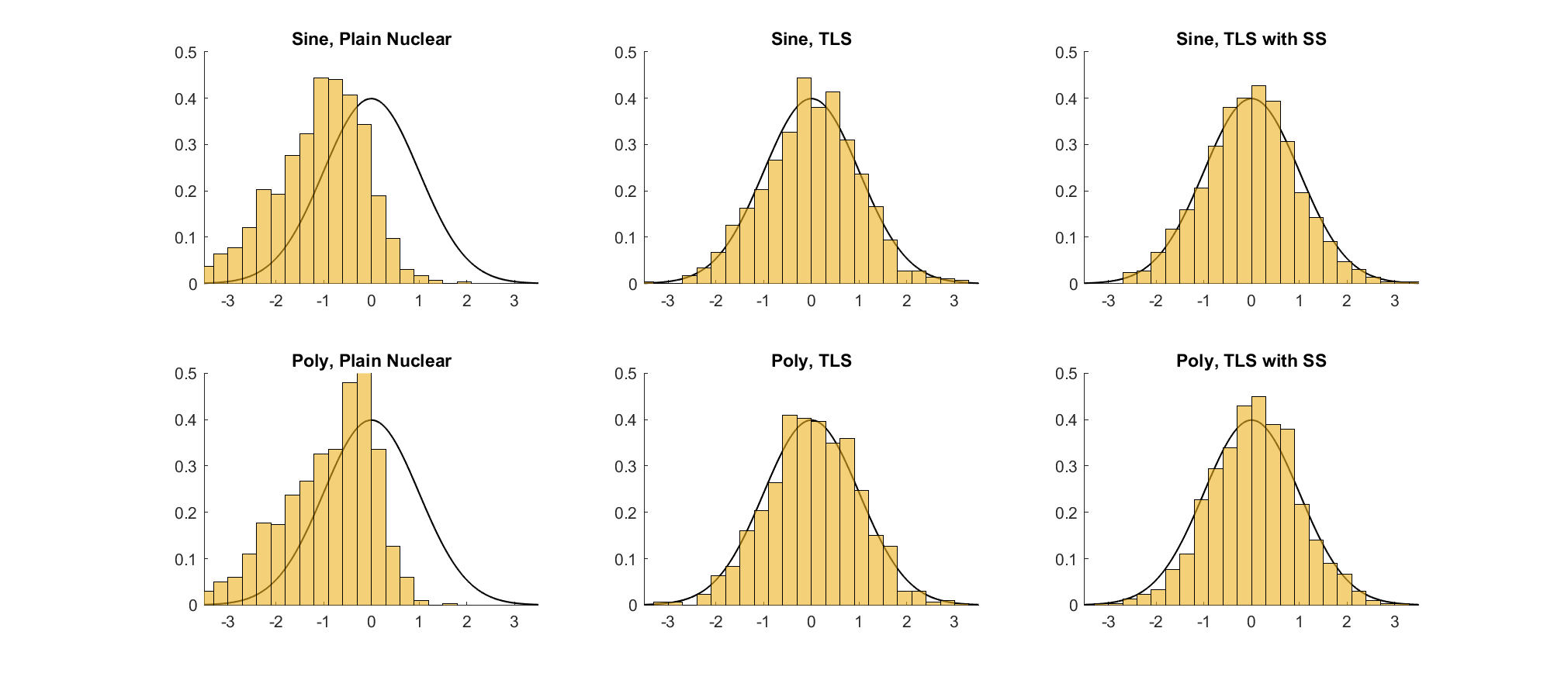} 
	\end{center}
	\vspace{-5mm}
		\caption{\small  Histograms of standardized estimates, $(   \widehat{M}_{it} - M_{it}) /se(   \widehat{M}_{it} )$}
        \label{fig:histocltcompare}
	\end{figure}
	
	Figure \ref{fig:histocltcompare} plots the scaled histograms of the standardized estimates with the standard normal density. As we expected in theory, it shows that the standardized TLS and the standardized TLS with SS fit the standard normal distribution well, while the standardized Plain Nuclear is biased. Without sample splitting, the TLS itself provides a good approximation to the standard normal distribution so that it can be used for the inference successfully. The coverage probabilities of confidence interval in Appendix also show similar results.

\begin{figure}[htb!]
	\begin{center}
        \includegraphics[width=0.8\linewidth]{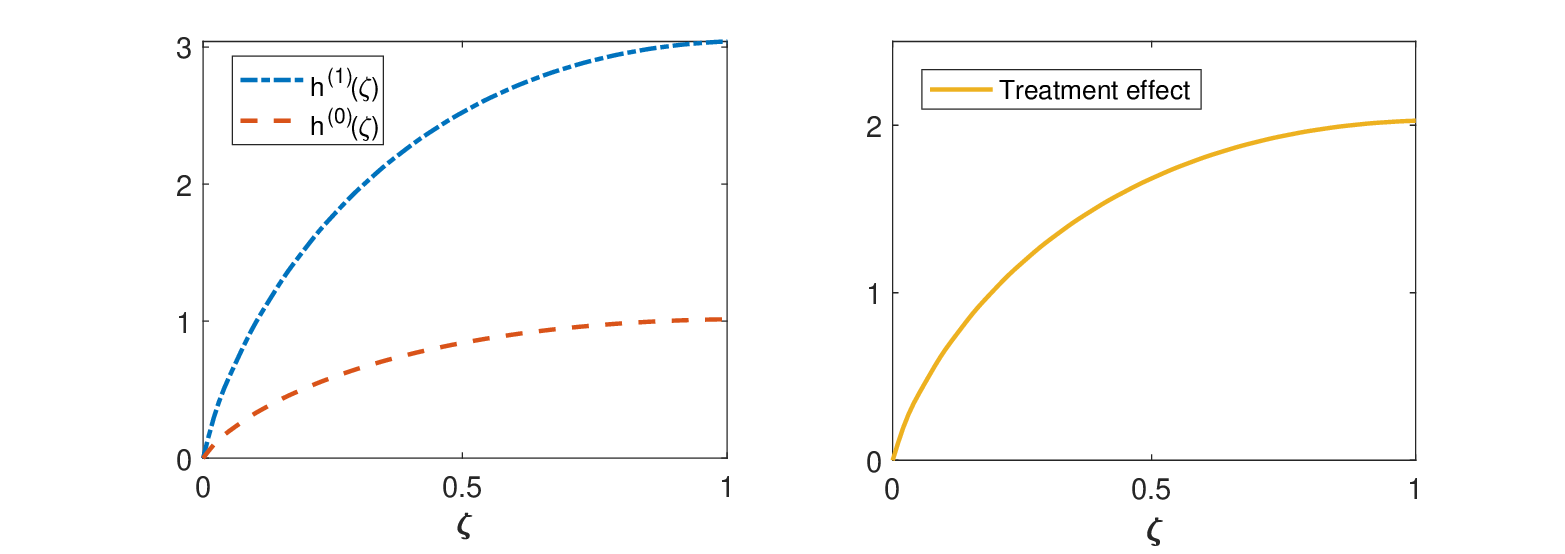} 	\vspace{-0.5cm}
	\end{center}
		\caption{\small  Shape of function $h_t^{(\iota)}(\zeta)$ and treatment effect function ($U_{t,r}=1$, $a=2$)}
		\label{fig:treatmentshape}
	\end{figure}

	Next, we study the finite sample performance of the average treatment effect estimator. Following Section \ref{sec:treatment}, for each $\iota \in \{0,1\}$, we generate the data from $y_{it}^{( \iota)}  = h_t^{(\iota)}(\zeta_{i})  + \varepsilon_{it}$, where $ h_t^{(0)}(\zeta) = \sum_{r=1}^{\infty} |U_{t,r}| r^{-a}\sin(r\zeta)$, $h_t^{(1)}(\zeta) = \sum_{r=1}^{\infty} (|U_{t,r}|+2)r^{-a}\sin(r\zeta)$. The power parameter $a>1$ controls the decay speed of the sieve coefficients. The forms of the above functions and the treatment effect $\Gamma_{it}= h_t^{(1)}(\zeta_i) -  h_t^{(0)}(\zeta_i)$ are in Figure \ref{fig:treatmentshape}.
	
	Here, $\varepsilon_{it}$ and $U_{t,r}$ are independently generated from the standard normal distribution and $\zeta_{i}$ is independently generated from $\text{Uniform}[0,1]$. The treatment pattern follows $\Upsilon_{it}\sim \text{Bernoulli}( p_i^{(1)}) $ and $p_i^{(1)} \sim \text{Uniform}[0.3,0.7]$.
	
	\begin{figure}[htb!]
	\begin{center}
	   \includegraphics[width=0.9\linewidth]{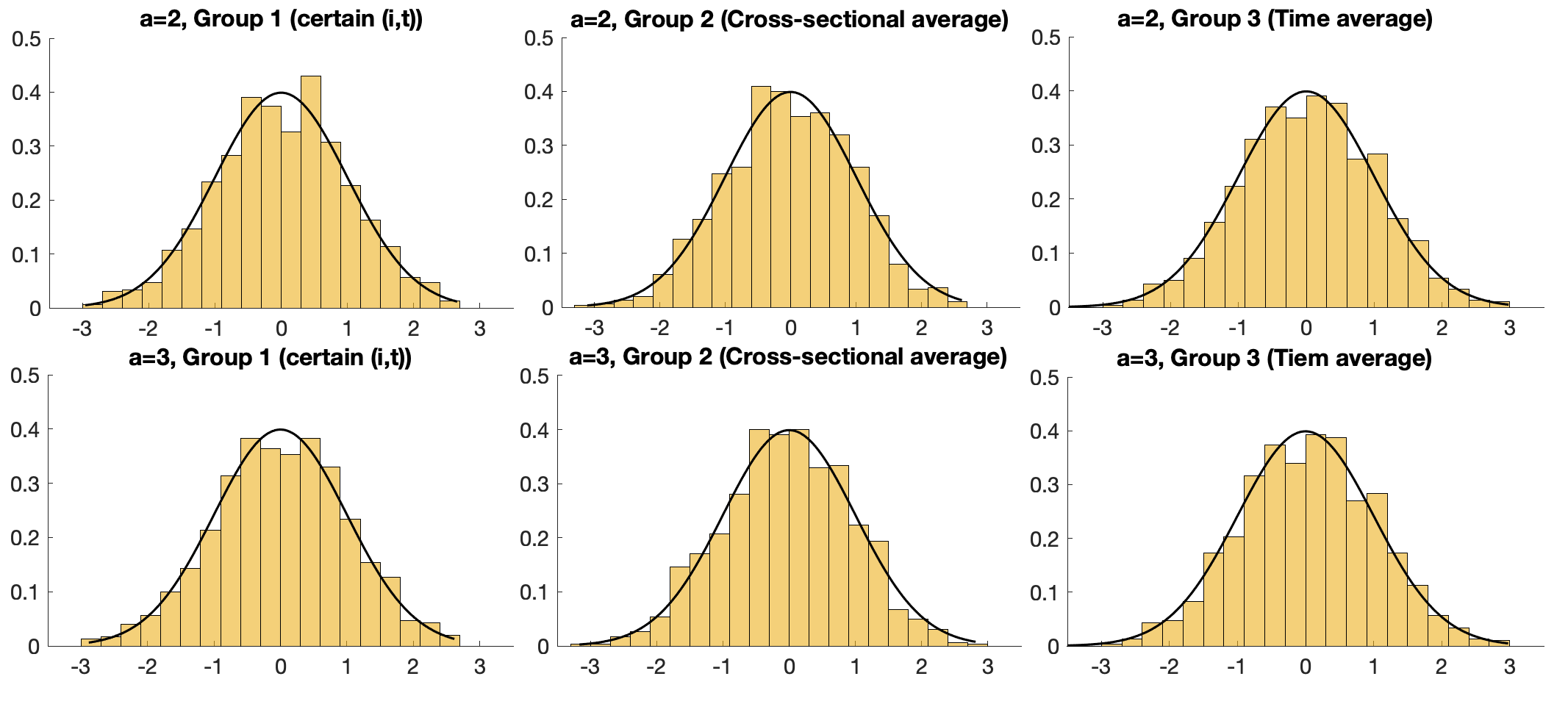} \vspace{-0.5cm}
	\end{center}
			\caption{\small  Histograms of standardized estimates, $ \frac{\frac{1}{|\calG|_o}\sum_{(i,t)\in\calG}  \widehat{\Gamma}_{it}  - \frac{1}{|\calG|_o}\sum_{(i,t)\in\calG}  \Gamma_{it} }{se\left(   \frac{1}{|\calG|_o}\sum_{(i,t)\in\calG}  \widehat{\Gamma}_{it}  \right) }$}		\label{fig:histotreatment}
		\centering
		\vspace{5mm}
		\begin{minipage}{1\textwidth}
			{\footnotesize NOTE:  Here, the sample size is $N=T=300$. \textquotedblleft Group 1\textquotedblright{} refers to $\calG_{1}$, \textquotedblleft Group 2\textquotedblright{} denotes $\calG_{2}$ and \textquotedblleft Group 3\textquotedblright{} refers to $\calG_{3}$.}
		\end{minipage}
	\end{figure}
	
	Figure \ref{fig:histotreatment} presents the scaled histograms of the standardized estimates of the average treatment effect estimators for the groups $\calG_1 = \{(i,t)\}$, $\calG_2 = \{(j,t):1\leq  j \leq N \}$, and $\calG_3 = \{(i,s): 1\leq  s  \leq T\}$. Here, the standard estimates are given as $$\frac{\frac{1}{|\calG|_o}\sum_{(i,t)\in\calG}  \widehat{\Gamma}_{it}  - \frac{1}{|\calG|_o}\sum_{(i,t)\in\calG} \Gamma_{it} }{se\left(   \frac{1}{|\calG|_o}\sum_{(i,t)\in\calG}  \widehat{\Gamma}_{it}  \right) }.$$ As expected in theory, the standardized estimates of the average treatment effect estimators of all groups approximately show the standard normal distribution. In addition, the coverage probabilities of the confidence interval in Appendix also show similar results.

\section{Conclusion}\label{sec:conclusion}
	This paper studies the inferential theory for low-rank matrices and provides an inference method for the average treatment effect as an application. Without the aid of sample splitting, our estimation procedure successfully resolves the problem of the shrinkage bias, and the resulting estimator attains the asymptotic normality. Unlike \cite{chernozhukov2019inference, chernozhukov2021inference} which exploit sample splitting, our estimation step is simple, and we can avoid some undesirable properties of sample splitting. In addition, this paper allows the heterogeneous observation probability and uses inverse probability weighting to control the effect of the heterogeneous observation probability.

 \section{Supplement Materials}
  For the sake of brevity, some of the technical proofs are relegated to the Supplement.

\appendix

{\LARGE 
\begin{center}
    APPENDIX
\end{center}
}

\section{Data-driven ways of choosing $K$} \label{sec:choosingK}

\noindent \textbf{Using a consistent estimator of $K$}

To choose the sieve dimension $K$, we can use the following rank estimator of $M^{\star}$ in the general approximate factor model $\widehat{K} = \sum_{r} 1\{\psi_r(\widetilde{M})\geq ((N+T)/2)^{\frac{11}{20}}\|\widetilde{M}\|^{\frac{1}{4}}\}$ where $\psi_{r}(\widetilde{M})$ denotes the $r$th largest singular value of $\widetilde{M}$. As noted in Claim F.1 (iii), it works as a consistent rank estimator for $M^{\star}$ in the general approximate factor model. By the same token in Footnote 5 of \cite{bai2003inferential}, our inferential theory for the general approximate factor model is not affected even if the rank $K$ is unknown and estimated using this estimator since $P( \widehat{K} = K  )\rightarrow 1$.

\noindent \textbf{Cross-validation method}

When the matrix of interest $M$ is approximated by a low-rank structure via a sieve representation like our main model, we can treat the sieve dimension $K$ as a tuning parameter. Hence, we introduce one data-driven way of selecting $K$ which exploits the cross-validation which is similar to the idea in \cite{athey2021matrix}. From the observed sample $\{(i,t) : \omega_{it}=1\}$, we randomly create a subsample by using a Bernoulli process, namely the subsample is $\{(i,t) : \omega_{it}  X_{it} = 1\}$ where $\{X_{it}\}_{i \leq N, t \leq T}$ are independent Bernoulli random variables of probability $\sum_{i,t} \omega_{it} /NT$, which is independent of $\{\omega_{it}\}_{i \leq N, t \leq T}$. This guarantees that we have $\sum_{i,t}\omega_{it} / NT \approx \sum_{i,t}\omega_{it}X_{it} / \sum_{i,t} \omega_{it}$. We then pre-specify the set of candidates of $K$ as $\{K_1, K_2, \ldots \}$ and compute the estimates $\widehat{M}_{K_1}, \widehat{M}_{K_2}, \ldots$, respectively, using only the subsample. To compare their out-of-sample performance, we measure the mean squared error of them on $\{(i,t) : \omega_{it}(1-X_{it})=1\}$. For robustness, we repeat this process five times, creating different independent subsamples each time, to obtain five mean squared errors for each $K \in \{K_1, K_2, \ldots \}$. The sieve dimension which minimizes the sum of five mean squared errors is chosen. In our simulation study, we use this method with $\{2,4,6,8,10\}$ as the set of candidates of $K$.

\section{Finite sample convergence rate}\label{sec:convergencerate}

For completeness, this section studies the finite sample convergence rate of our estimator. First, we provide several conditions. Here, $a \lesssim b$ means $|a|/|b| \leq C$ for some constant $C>0$. $a \ll b$ indicates $|a| \leq c|b|$ for some sufficiently small constant $c>0$.

\begin{assumption}[Sieve representation]\label{asp:nonparametric_sieve_finite}
		\textit{(i) $\{h_{t}(\cdot) \}_{t\leq T}$ belong to ball $\calH\left( \calZ,\norm{\cdot }_{L_2},C  \right)$ inside a Hilbert space spanned by the basis $\{\phi_r \}_{r\geq 1}$, with a uniform $L_2$-bound $C$: 
			$ \sup_{h\in \calH( \calZ,\norm{\cdot}_{L_2})}\|h\| \leq C,$ where $\calZ$ is the support of $\zeta_{i}$.\\
			(ii) The sieve approximation error satisfies: For some $\nu>0$, $\max_{i,t}|M^{R}_{it}| \leq C K^{-\nu}$.\\
			(iii) For some $C>0$, $\max_{r \leq K}\sup_{\zeta} |\phi_r (\zeta)| < C$. In addition, there is $\eta > 0$ such that $\psi_{\min}^{-1}\left( S_{\beta} \right) < \eta$ and $\psi_{\min}^{-1}\left( S_F \right) < \eta$. \\
			(iv) $ \sum_{i,t} h_t^2 ( \zeta_{i}) \lesssim NT$.\\
			(v) There are constants $\delta,g \geq 0$ such that
			$\psi_{1}(Q) / \psi_{K}(Q) \lesssim K^{\delta}$, $\min_{1\leq r \leq K-1} \psi_{r}(Q) - \psi_{r+1}(Q) \geq c K^{-g}$ for some constant $c>0$.
			}
	\end{assumption}

This condition is basically the same as Assumption \ref{asp:nonparametric_sieve}, and we modify some notation to be suitable for finite sample analysis.

\begin{assumption}[Parameter size and signal-to-noise ratio]\label{asp:nonparametric_parameters_finite}
		\textit{Let $\gamma = \frac{p_{\max}}{p_{\min}}$ and $\tilde{\vartheta} = \max\{\vartheta, \log N + \log T\}$. Then, we have
			\begin{align*}     
				&(i)\ \ \tilde{\theta}\eta^{\frac{3}{2}} \gamma^{\frac{5}{2}}K^{(2+2g+\frac{9}{2}\delta)}\max\{\sqrt{N\log N},\sqrt{ T\log T} \} \ll p_{\min}^{\frac{1}{2}} \min\{N,T\},\qquad\qquad\qquad\qquad\qquad\qquad\\
			 &\qquad \gamma^{\frac{3}{2}}K^{(g+\frac{3}{2}\delta)} \max\{N,T\} \ll p_{\min}^{\frac{1}{2}} \min\{\sqrt{N\log N},\sqrt{ T\log T} \} \psi_{NT}, \\
				&(ii)\ \ 
				\min\{|\calI|_o^{\frac{1}{2}}, |\calT|_o^{\frac{1}{2}} \}\max\{\sqrt{N},\sqrt{T} \}  \ll p_{\min}^{\frac{1}{2}} K^{(\nu - \frac{1}{2} - 2\delta)} ,\\
            &\qquad\min\{|\calI|_o^{\frac{1}{2}}, |\calT|_o^{\frac{1}{2}} \}\max\{\sqrt{N},\sqrt{T} \} \sqrt{NT} \ll \gamma^{\frac{1}{2}} \psi_{NT} K^v .
			\end{align*}}
\end{assumption}

The above condition is weaker than the condition for the asymptotic normality (Assumption \ref{asp:nonparametric_parameters}). For example, Assumption \ref{asp:nonparametric_parameters_finite} (i) does not restrict the size of the interesting group, $\min\{|\calI|_o,|\calT|_o \}$, unlike Assumption \ref{asp:nonparametric_parameters} (i). Hence, we can deal with the case where $|\calI|_o = N$ and $|\calT|_o = T$. In addition, it allows for a weaker signal-to-noise ratio than that of Assumption \ref{asp:nonparametric_parameters}.

\begin{proposition}\label{pro:convergencerate}
 Suppose Assumptions \ref{asp:nonparametric_dgp}, \ref{asp:clusterdependence}, \ref{asp:nonparametric_sieve_finite}, and \ref{asp:nonparametric_parameters_finite}. Then, with probability at least $1 - O(\min\{N^{-3},T^{-3}\})$, we have
\begin{align*}
\norm{\frac{1}{|\calG|_o} \sum_{(i,t)\in \calG}\widehat{M}_{it} - \frac{1}{|\calG|_o} \sum_{(i,t)\in \calG} M_{it}} 
&\leq C \left( \frac{\sigma \eta^{\frac{1}{2}}K^{\frac{1}{2}} \max\{\sqrt{\log N}, \sqrt{\log T} \}}{p_{\min}^{\frac{1}{2}} \sqrt{N |\calT|_o}, } + \frac{\sigma \eta^{\frac{1}{2}}K^{\frac{1}{2}} \max\{\sqrt{\log N}, \sqrt{\log T} \}}{p_{\min}^{\frac{1}{2}} \sqrt{T |\calI|_o} } \right.\\
& \quad \left. + \frac{\sigma \tilde{\vartheta} \gamma^{\frac{7}{2}} K^{(4 + 2g + \frac{13}{2}\delta )}\eta^{3}\max\{\log N,\log T\}}{p_{\min}^{\frac{3}{2}}\min\{N,T\}}+ \frac{\sigma^3 \gamma^2 K^{(\frac{7}{2}\delta + g + 1)}\eta^{\frac{1}{2}}\max\{N,T\}}{p_{\min}^2 \psi_{NT}^2} \right)
\end{align*}
for some constant $C>0$.   
\end{proposition}

The first two terms represent the asymptotically normal distribution parts, while the last two terms are the residual parts related to the estimation errors of $\beta_i$ and $f_t$. If we ignore some small parameters and logarithmic terms, the convergence rate of the first two terms is reduced to
$$\frac{1}{\sqrt{N|\calT|_o}} + \frac{1}{\sqrt{T|\calI|_o}}.$$
However, if both $|\calI|_o$ and $|\calT|_o$ are large, as in the case where $|\calI|_o=N$ and $|\calT|_o=T$, the asymptotically normal parts cannot dominate the residual parts. Thus, we are unable to derive the inferential theory in this case. For inference, at least one part of the asymptotically normal terms should dominate other residual terms. On the other hand, in terms of the convergence rate, the large sizes of $|\calI|_o$ and $|\calT|_o$ are beneficial.

\section{Inferential theory for the general approximated factor model} \label{sec:factor_assumption}

This section provides assumptions for the asymptotic normality of the estimator of the group average of $M_{it}$ for the general approximated factor model having the form $Y = M + \calE$ where $M = M^\star + M^R$, $rank(M^\star) = r$. For this, we define some additional notations. The condition number of $M^{\star}$ is defined as $q \coloneqq \psi_{\max}(M^{\star}) / \psi_{\min}(M^{\star})$. Define $\bar{c}= \min_{1\leq r \leq K+1} \abs{c_{r-1}^2 -  c_{r}^2}$, where $c_r \coloneqq \psi_{r}(M^{\star})/\psi_{\min}(M^{\star})$, and $c_{\inv}\coloneqq 1/\bar{c}$.\footnote{We set $c_0\coloneqq\infty$. Note that $\psi_{r}=0$ for $r >K$, and that $c_1^2 = q^2 \geq c_r^2 \geq c_K^2 = 1$ for all $1\leq r \leq K$. $\bar{c}$ is always smaller than $1$ since $c_K ^2 - c_{K+1}^2 = 1$. Hence, $c_{\inv} \geq 1$. We allow $c_{\inv}$ to increase slowly as $N$ and $T$ increase.} 

\begin{assumption}[Incoherence]\label{asp:incoherence}
	\textit{The matrix $M^{\star}$ satisfies $\mu$-incoherence condition. That is, 
	$\norm{U_{M^{\star}}}_{2,\infty} \leq \sqrt{\frac{\mu}{N}}\norm{U_{M^{\star}}}_F=\sqrt{\frac{\mu K}{N}}$ and $\norm{V_{M^{\star}}}_{2,\infty} \leq \sqrt{\frac{\mu}{T}}\norm{V_{M^{\star}}}_F=\sqrt{\frac{\mu K}{T}}$ with probability converging to 1. Here, $\mu$ is allowed to increase as $N,T$ increase.}
\end{assumption}

\begin{assumption}[Parameters size]\label{asp:smallop1}
	\textit{Let $\gamma = \frac{p_{\max}}{p_{\min}}$ and $\tilde{\vartheta} = \max \{\vartheta , \log N + \log T \}$. Then, we have
	\begin{align*}          
			&\text{(i)   }\min\{|\calI|_o^{1/2},|\calT|_o^{1/2}\} \tilde{\vartheta} c_{\inv} q^{\frac{15}{2}}\mu^3 K^{4}\gamma^{\frac{7}{2}} \max\{\sqrt{N \log N},\sqrt{ T \log T} \} = o_P(p_{\min}\min\{N,T\}),\quad \quad \quad \quad \quad  \quad \quad \quad \quad \quad \quad \\
			&\text{(ii)   }\min\{|\calI|_o^{1/2},|\calT|_o^{1/2}\} \tilde{\vartheta} c_{\inv}^2 q^{7}\mu^{\frac{5}{2}} K^{\frac{7}{2}} \gamma^{4}\max\{N \sqrt{\log N}, T\sqrt{\log T} \} = o_P(\psi_{\min} p_{\min}^{\frac{3}{2}} \min\{\sqrt{N},\sqrt{T}\}),\\
			&\text{(iii)   }\min\{|\calI|_o^{1/2},|\calT|_o^{1/2}\} \vartheta c_{\inv}^2 q^{6}\mu^{2} K^{\frac{7}{2}}  \gamma^{\frac{7}{2}}\max\{N^{\frac{3}{2}} \sqrt{\log N}, T^{\frac{3}{2}} \sqrt{\log T} \} = o_P(\psi^2_{\min} p_{\min} ),\\
			&\text{(iv)   }\min\{|\calI|_o^{1/2},|\calT|_o^{1/2}\} c_{\inv} q^{\frac{7}{2}}\mu^{\frac{1}{2}} K \gamma^{3}\max\{N^{2}, T^{2} \}\min\{\sqrt{N}, \sqrt{T}\} = o_P(\psi^3_{\min} p^{\frac{3}{2}}_{\min} ).
		\end{align*}
		}
\end{assumption}

\begin{assumption}[Low-rank approximation error $M^{R}$]\label{asp:lowrankapprx}
	\textit{The low-rank approximation error $M^{R}$ satisfies the following condition:
	\begin{align*}
	    	\max_{i,t}\abs{M^{R}_{it}} 
	    	= & o_P\left( \frac{p_{\min}^{\frac{5}{2}}}{\min\{|\calI|_o^{1/2},|\calT|_o^{1/2}\}p_{\max}^{2}q^2\mu^{\frac{3}{2}} K^{\frac{3}{2}} \max\{\sqrt{N},\sqrt{ T} \}} \right. \\
	    	&\left. + \frac{\psi_{\min}p_{\min}^2}{\min\{|\calI|_o^{1/2},|\calT|_o^{1/2}\}p_{\max}^{\frac{3}{2}}q\mu^{\frac{1}{2}} K^{\frac{1}{2}} \max\{\sqrt{N},\sqrt{ T} \}\sqrt{NT}}  \right)
	\end{align*}
	}
\end{assumption}
Then, the estimator for the group average of $M_{it}$ has the asymptotic normality as follows. 

\begin{theorem} \label{thm:generalfactor_groupclt}
Suppose Assumptions \ref{asp:nonparametric_dgp}, \ref{asp:clusterdependence} and \ref{asp:incoherence}-\ref{asp:lowrankapprx} hold. In addition, suppose that\\ $\norm{\frac{\sqrt{N}}{|\calI|_o}\sum_{i \in \calI}U_{M^*,i}} \geq c$ and $\norm{\frac{\sqrt{T}}{|\calT|_o}\sum_{t \in \calT}V_{M^*,t}} \geq c$ for some constant $c>0$. Then, 
$$\calV_{\calG}^{-\frac{1}{2}}\left( \frac{1}{|\calG|_o}\sum_{(i,t)\in\calG}  \widehat{M}_{it} - \frac{1}{|\calG|_o}\sum_{(i,t)\in\calG}  M_{it} \right) \conD \calN(0,1),$$
\begin{align*}
  \text{where  }\ \   \calV_{\calG}&=\frac{1}{|\calT|_o^2} \sum_{t\in\calT}\bar{\beta}_{\calI}^{\prime}\left( \sum_{j=1}^{N} \omega_{jt}\beta_{j}\beta_{j}^{\prime} \right)^{-1}\left( \sum_{j=1}^{N} \omega_{jt} \sigma^2_{jt} \beta_{j}\beta_{j}^{\prime} \right) \left( \sum_{j=1}^{N} \omega_{jt}\beta_{j}\beta_{j}^{\prime} \right)^{-1}\bar{\beta}_{\calI} \\
    & \ \ + \frac{1}{|\calI|_o^2} \sum_{i\in\calI}\bar{F}_\calT^{\prime} \left( \sum_{s=1}^{T}\omega_{is}F_s F_s^{\prime}  \right)^{-1}\left( \sum_{s=1}^{T}\omega_{is}\sigma^2_{is}F_s F_s^{\prime}  \right) \left( \sum_{s=1}^{T}\omega_{is}F_s F_s^{\prime}  \right)^{-1}\bar{F}_\calT , \qquad \qquad \qquad \qquad \qquad 
\end{align*}
$\bar{\beta}_{\calI} = \frac{1}{|\calI|_o}\sum_{i \in \calI}\beta_{i}$, $\bar{F}_\calT = \frac{1}{|\calT|_o}\sum_{s \in \calT}F_{s}$. In addition, Assumptions \ref{asp:incoherence} - \ref{asp:lowrankapprx} are satisfied under Assumptions \ref{asp:nonparametric_sieve} - \ref{asp:nonparametric_parameters} by setting $\mu = C\eta$ for some constant $C>0$. 
\end{theorem}

In fact, Assumptions \ref{asp:incoherence} - \ref{asp:lowrankapprx} are verified by Lemma F.1.

\begin{theorem} [Feasible CLT]\label{thm:generalfactor_feasibleclt}
Under the assumptions of Theorem \ref{thm:generalfactor_groupclt}, we have
$$\widehat{\calV}_{\calG}^{-\frac{1}{2}}\left( \frac{1}{|\calG|_o}\sum_{(i,t)\in\calG}  \widehat{M}_{it}  - \frac{1}{|\calG|_o}\sum_{(i,t)\in\calG}  M_{it} \right) \conD \calN(0,1),$$ 
where $\widehat{\calV}_\calG$ is the same as the one in Theorem \ref{thm:feasibleclt}.
\end{theorem}

\section{Formal definitions of the non-convex estimator and the leave-one-out estimator} \label{sec:estimator_definition}

	Here, we introduce formal definitions of the non-convex optimization estimator $( \widetilde{W}^{[l]},\widetilde{Z}^{[l]})$ and the leave-one-out estimator $( \breve{W}^{(l)},\breve{Z} ^{(l)}) $ where $1 \leq l \leq N+T$. We start with defining the following two loss functions:
\begin{align}
	&\label{eq:infsob}  f^{\infs}(w,z)   \coloneqq  \frac{1}{2}\|\Pi^{-\frac{1}{2}}\calP_{\Omega}\left( wz^{\prime} - Y \right)\|_F^2 + \frac{\lambda}{2}\|w\|_F^2 + \frac{\lambda}{2}\|z\|_F^2, \\
	&\label{eq:looob}  f^{\infs,(l)}(w,z)  \\
	&\nonumber\coloneqq \begin{cases}
		\frac{1}{2} \norm{\Pi^{-1/2} \mathcal{P}_{\Omega_{-l,\cdot}}(wz'-Y)}_F^2 + \frac{1}{2} \norm{\mathcal{P}_{l,\cdot}(wz'-M^{\star})}_F^2  + \frac{\lambda}{2}\norm{w}_F^2+\frac{\lambda}{2}\norm{z}_F^2, \quad \text{if $1 \leq l \leq N,$} \\
		\frac{1}{2} \norm{\Pi^{-1/2} \mathcal{P}_{\Omega_{\cdot, -(l-N)}}(wz'-Y)}_F^2 + \frac{1}{2} \norm{\mathcal{P}_{\cdot, (l-N)}(wz'-M^{\star})}_F^2  + \frac{\lambda}{2}\norm{w}_F^2+\frac{\lambda}{2}\norm{z}_F^2,  \end{cases}\\
	& \nonumber\quad \quad \quad \quad \quad \quad \quad \quad \quad \quad \quad \quad \quad \quad \quad \quad \quad \quad \quad \quad \quad \quad \quad \quad \quad \quad \quad \quad \quad \quad \quad \text{if $N+1 \leq l \leq N+T$,}\end{align}%
where $w$ and $z$ are $N \times K$ and $T \times K$ matrices, respectively. The loss function \eqref{eq:infsob} is for the non-convex optimization estimator $( \widetilde{W}^{[l]},\widetilde{Z}^{[l]} )$ and the loss function \eqref{eq:looob} is for the leave-one-out estimator $\breve{W}^{(l)}$. In the loss function \eqref{eq:looob}, we use the following definitions. Let $\calC_{g(i)}$ be the cluster where the unit $i$ is included in. For each $N \times T$ matrix $D$, let $\calP_{\Omega}= \Omega \circ D$. Also, for each $N \times T$ matrix $D$ and for each $1 \leq l \leq N$, let $\mathcal{P}_{\Omega_{-l,\cdot}}(D)\coloneqq \Omega_{-l,\cdot}\circ D$ where $\Omega_{-l,\cdot}\coloneqq[\omega_{js}1\{j \notin \calC_{g(l)}\}]_{N \times T}$, and $\mathcal{P}_{l,\cdot}(D)\coloneqq E_{l,\cdot} \circ D$ where $E_{l,\cdot} \coloneqq [1\{j \in \calC_{g(l)}\}]_{N \times T}$. Roughly speaking, $ f^{\infs,(l)}$ changes $\{p_j^{-1}\omega_{js},y_{js}\}_{j \in \calC_{g(l)}, s \leq T}$ in $ f^{\infs}$ to its (approximate) population mean $\{1,M_{js}^{\star}\}_{j \in \calC_{g(l)}, s \leq T}$. Hence, the leave-one-out estimator constructed from the loss function $ f^{\infs,(l)}$ can be independent of $\{\omega_{ls}, \varepsilon_{ls}\}_{s\leq T}$ because $ f^{\infs,(l)}$ excludes $\{\omega_{js}, \varepsilon_{js}\}_{j \in \calC_{g(l)},s\leq T}$ which is in the cluster where the unit $l$ is included in. 

On the other hand, for each $N+1 \leq l \leq N+T$, we define $\mathcal{P}_{\Omega_{\cdot, -(l-N)}}(D)\coloneqq \Omega_{\cdot, -(l-N)}\circ D$ where $\Omega_{\cdot, -(l-N)}\coloneqq[\omega_{js}1\{s \neq l-N\}]_{N \times T}$, and $\mathcal{P}_{\cdot, (l-N)}(D)\coloneqq E_{\cdot, (l-N)} \circ D$ where $E_{\cdot, (l-N)} \coloneqq [1\{s=l-N\}]_{N \times T}$. In this case, $ f^{\infs,(l)}$ changes $\{p_j^{-1}\omega_{js}, y_{js}\}_{j \leq N, s = l-N}$ in $ f^{\infs}$ to $\{1, M_{js}^{\star }\}_{j \leq N, s = l-N}$. So, the leave-one-out estimator constructed from $ f^{\infs,(l)}$ is independent of $\{\omega_{j,(l-N)}, \varepsilon_{j,(l-N)}\}_{j\leq N}$ because $ f^{\infs,(l)}$ excludes $\{\omega_{j,(l-N)}, \varepsilon_{j,(l-N)}\}_{j\leq N}$ and $\omega_{js}$, $\varepsilon_{js}$ are independent across time.

To define the gradient descent iterates, we denote the singular value decomposition (SVD) of $M^{\star}$ by $U_{M^{\star}} D_{M^{\star}} V_{M^{\star}}'$ where $U_{M^{\star}}'U_{M^{\star}}=V_{M^{\star}}'V_{M^{\star}}=I_K$. $D_{M^{\star}}$ is a $K \times K$ diagonal matrix with singular values in descending order, i.e., $D_{M^{\star}}=\diag(\psi_{1},\dots,\psi_{K})$ where $\psi_{\max} = \psi_{1} > \cdots > \psi_{K} = \psi_{\min}>0$. Then, based on \eqref{eq:infsob}, we define the following gradient descent iterates: 
\begin{align}\label{alg:nonconvex}
	\begin{bmatrix}
		W^{\tau+1} \\ Z^{\tau+1}
	\end{bmatrix}
	=\begin{bmatrix}
		W^{\tau} - \eta \nabla_W f^{\infs}(W^{\tau}, Z^{\tau}) \\ Z^{\tau} - \eta \nabla_Z f^{\infs}(W^{\tau}, Z^{\tau})
	\end{bmatrix}
\end{align}
where $W^0=W \coloneqq U_{M^{\star}}D_{M^{\star}}^{\frac{1}{2}}$, $Z^0=Z \coloneqq V_{M^{\star}}D_{M^{\star}}^{\frac{1}{2}}$, $\tau=0,1, \ldots, \tau_0-1$, and $\tau_0=\max\{N^{18}, T^{18}\}$. Here, $\eta>0$ is the step size. Similarly, for \eqref{eq:looob}, we define
\begin{align}\label{alg:loo}
	\begin{bmatrix}
		W^{\tau+1, (l)} \\
		Z^{\tau+1, (l)}
	\end{bmatrix}
	= \begin{bmatrix}
		W^{\tau,(l)} - \eta \nabla_W f^{\infs,(l)}(W^{\tau,(l)}, Z^{\tau,(l)}) \\
		Z^{\tau,(l)} - \eta \nabla_Z f^{\infs,(l)}(W^{\tau,(l)}, Z^{\tau,(l)})
	\end{bmatrix}
\end{align}%
where $W^{0,(l)}=W$, $Z^{0,(l)}=Z$. Note that the gradient descent iterates in \eqref{alg:nonconvex} and \eqref{alg:loo} cannot be feasibly computed because the initial values ($W$, $Z$), the missing probability ($\Pi$), and the cluster structure are unknown. However, it does not cause any problem in the paper since we do not need to actually compute $W^{\tau}, Z^{\tau}, W^{\tau, (l)}$, and $Z^{\tau, (l)}$ and only use their existence and theoretical properties for the proof. We also define for each $\tau$ and $l$,
\begin{align*}
	&H^{\tau} \coloneqq \argmin_{O\in \calO^{K \times K}}\norm{
		\calF^{\tau} 
		O - 
		\calF}_F ,  \ \ \
	H^{\tau, (l)} \coloneqq \argmin_{O \in \mathcal{O}^{K \times K}}\norm{
		\calF^{\tau, (l)}
		O - 
		\calF}_F,\\
	&Q^{\tau, (l)} \coloneqq \argmin_{O \in \mathcal{O}^{K \times K}}\norm{
		\calF^{\tau, (l)}
		O - 
		\calF^{\tau} H^{\tau}}_F,\ \ \text{where } 
	\calF^{\tau} \coloneqq 
	\begin{bmatrix}
		W^{\tau} \\
		Z^{\tau}
	\end{bmatrix},\ \
	\calF^{\tau, (l)}\coloneqq 
	\begin{bmatrix}
		W^{\tau,(l)} \\
		Z^{\tau,(l)}
	\end{bmatrix}, \ \
	\calF \coloneqq 
	\begin{bmatrix}
		W \\
		Z
	\end{bmatrix},
\end{align*} 
and $\calO^{K \times K}$ is the set of $K \times K$ orthogonal matrix. Importantly, by the definition, $H^{\tau, (l)}$ is also independent to the observations in $l$. 

In this paper, as emphasized in the main text, we consider the non-convex optimization estimator $( \widetilde{W}^{[l]},\widetilde{Z}^{[l]} )$ and the leave-one-out estimator $( \breve{W}^{(l)},\breve{Z} ^{(l)}) $ at two different stopping points. Let $\tau_l^* \coloneqq \argmin_{0\leq \tau < \tau_o}\norm{\nabla f^{\infs,(l)}(W^{\tau,(l)},Z^{\tau,(l)})}_F$. First, we use the stopping point $\tau^*_l$, i.e.,
$$         
( \widetilde{W}^{[l]},\widetilde{Z}^{[l]} ) \coloneqq (W^{\tau_l^*},Z^{\tau_l^*}) \quad \text{from \eqref{alg:nonconvex}}, \quad
( \breve{W}^{(l)},\breve{Z} ^{(l)}) \coloneqq (W^{\tau_l^*,(l)},Z^{\tau_l^*,(l)}) \quad \text{from \eqref{alg:loo}},$$
and $\widetilde{H}^{[l]} \coloneqq H^{\tau_l^{*}}$, $\breve{H}^{(l)}\coloneqq H^{\tau_l^{*},(l)}$. For each $l$, we set the same iteration number $\tau_l^*$ for the non-convex optimization estimator $( \widetilde{W}^{[l]},\widetilde{Z}^{[l]} )$ and the leave-one-out estimator $( \breve{W}^{(l)},\breve{Z} ^{(l)}) $ to ensure that they are close to each other. Note that, although the loss function \eqref{eq:infsob} does not depend on $l$, due to $\tau_l^*$, the non-convex optimization estimator $( \widetilde{W}^{[l]},\widetilde{Z}^{[l]} )$ depend on $l$. Namely, $( \widetilde{W}^{[l]},\widetilde{Z}^{[l]} )$ is selected to be close to the leave-one-out estimator $( \breve{W}^{(l)},\breve{Z} ^{(l)}) $ among many gradient descent iterates in \eqref{alg:nonconvex}. At last, we choose $H_4^{[l]}$ so that $\psi_{\min}^{-1/2}\widetilde{W}^{[l]} H_4^{[l]}$ is the left singular vector of $\widetilde{W}^{[l]} \widetilde{Z}^{[l]\prime}$.

Secondly, we use the stopping point  $\tau^* \coloneqq \argmin_{0\leq \tau < \tau_o}\norm{\nabla f^{\infs}(W^{\tau},Z^{\tau})}_F$. For brevity, we will use the same notations for the estimators. Namely, $$         
( \widetilde{W}^{[l]},\widetilde{Z}^{[l]} ) \coloneqq (W^{\tau^*},Z^{\tau^*}) \quad \text{from \eqref{alg:nonconvex}}, \quad
( \breve{W}^{(l)},\breve{Z} ^{(l)}) \coloneqq (W^{\tau^*,(l)},Z^{\tau^*,(l)}) \quad \text{from \eqref{alg:loo}},$$
and $\widetilde{H}^{[l]} \coloneqq H^{\tau^{*}}$, $\breve{H}^{(l)}\coloneqq H^{\tau^{*},(l)}$. Also, $H^{[l]}_4$ is defined similarly. Here, we are abusing notation in the sense that $( \widetilde{W}^{[l]},\widetilde{Z}^{[l]} ) $, $\widetilde{H}^{[l]}$ and $H^{[l]}_4$ do not actually depend on $l.$ However, this notational abuse is going to make the proofs more streamlined.

\begin{remark}\label{rem:simplenotation}
 In the main text, to facilitate understanding and save space, we use simpler notations. Specifically, $( \breve{\beta}^{\mathrm{full},t},\breve{\beta}^{(-t)},\breve{\beta}^{\{-i\}})$ in the main text is the same as
$$
\left( \breve{\beta}^{\mathrm{full},t},\breve{\beta}^{(-t)},\breve{\beta}^{\{-i\}}\right) \coloneqq \left(\sqrt{N}\widetilde{W}^{[N+t]}\widetilde{H}^{[N+t]}D_{M^{\star}}^{-\frac{1}{2}}, \sqrt{N}\breve{W}^{(N+t)}\breve{H}^{(N+t)}D_{M^{\star}}^{-\frac{1}{2}},\sqrt{N}\breve{W}^{(i)}\breve{H}^{(i)}D_{M^{\star}}^{-\frac{1}{2}}\right).
$$
\end{remark}

\section{Key part of proofs}

As we mentioned in Section \ref{subsec:debiasstrategys}, the key for having an unbiased estimator for $M_{it}$ is showing the following proposition:
\begin{proposition} \label{pro:estimatef}
	Suppose assumptions of Theorem \ref{thm:generalfactor_groupclt} hold.\footnote{By Lemma F.1, the assumptions of Theorem \ref{thm:generalfactor_groupclt} are satisfied under the assumptions of Theorem \ref{thm:generalfactor_groupclt}.} Then, there is a $K\times K$ matrix $H_2$ so that
\begin{align*}
		&\sqrt{N}(\widehat{F}_t - H_2 F_t) = \sqrt{N}H_2\left( \sum_{j=1}^{N}\omega_{jt}\beta_{j} \beta_{j}^{\prime}   \right)^{-1}\left( \sum_{j=1}^{N}\omega_{jt} \beta_{j} \varepsilon_{jt}  \right) + \sqrt{N} R^{F}_t,\\
&\max_{t}\|\sqrt{N} R^{F}_t\|\\
&=O_P\left(  
\frac{\sigma p_{\max}^{\frac{3}{2}}\vartheta c_{\inv}q^{\frac{11}{2}}\mu^{\frac{3}{2}}K^{\frac{5}{2}}\sqrt{N}\max\{\sqrt{\log N},\sqrt{\log T} \}}{ p_{\min}^3 \min \{N, T \}} 
+
\frac{\sigma^2 p_{\max}^{\frac{5}{2}}\vartheta c_{\inv}^2 q^{3} \mu K^{2} \sqrt{N}\max\{\sqrt{N\log N},\sqrt{T\log T} \}}{\psi_{\min} p_{\min}^4 \min \{\sqrt{N}, \sqrt{T} \}}
\right.\\
&\quad \quad \quad \left. + \frac{\sigma^3 p_{\max}^{\frac{3}{2}} c_{\inv} q^{\frac{5}{2}} K^{\frac{1}{2}}\sqrt{N} \max\{N,T \}}{ \psi_{\min}^2 p_{\min}^3 } + \frac{p_{\max}^{\frac{1}{2}}\sqrt{N}}{p_{\min}} \max_{it}\abs{M_{it}^R} \right) = o_P(1).
\end{align*}
\end{proposition}

 \subsection{Important Lemmas} \label{subsec:important_lemma}

An important step is to show that uniformly in $t$, the following two terms are negligible:
\begin{gather}\label{eqb.0ada}
\frac{1}{\sqrt{N}} \sum_{j=1}^N \omega_{jt}\varepsilon_{jt}(\widetilde{\beta}_j -\breve{\beta}^{\mathrm{full},t}_j ) ,\quad  \frac{1}{\sqrt{N}} \sum_{j=1}^N \omega_{jt}\varepsilon_{jt}( \breve{\beta}^{\mathrm{full},t}_j-    \breve{\beta}^{(-t)}_j).
\end{gather}
The proof follows from  Lemma \ref{lem:CCMFY} below.

\begin{lemma} \label{lem:CCMFY}

Suppose assumptions of Theorem \ref{thm:generalfactor_groupclt} hold. Uniformly in $t\leq T$, the two terms in (\ref{eqb.0ada}) are both  $o_P(1)$. Specifically, their order is 
$$
O_P\left( \frac{\sigma^2 p_{\max}^{\frac{3}{2}} \vartheta^{\frac{1}{2}} c_{\inv}q^{\frac{9}{2}}\mu^{\frac{1}{2}}K^{\frac{3}{2}}\sqrt{N}\max\{\sqrt{N\log N},\sqrt{T\log T}\}}{p_{\min}^2 \min\{\sqrt{N},\sqrt{T} \}\psi_{\min}}
+ \frac{\sigma^3 p_{\max}^{\frac{3}{2}} c_{\inv} q^{\frac{5}{2}} K^{\frac{1}{2}}\sqrt{N}\max\{N,T \}}{p_{\min}^2\psi_{\min}^2}\right).
$$
In addition, we have the following results:
\begin{align*}
		&(i) \ \ \max_{t} \|\widetilde{W}^{[t+N]}\widetilde{H}^{[t+N]} - \breve{W}^{(t+N)}\breve{H}^{(t+N)} \|_F = O_P\left( \frac{\sigma p_{\max}^{\frac{1}{2}}\vartheta^{\frac{1}{2}} q^{\frac{3}{2}}\mu^{\frac{1}{2}}K^{\frac{1}{2}}\max\{\sqrt{N\log N},\sqrt{T\log T}\}
		}{p_{\min}\psi_{\min}^{1/2} \min\{\sqrt{N},\sqrt{T} \} }\right) ,\\
		&(ii) \ \ \max_{t}\|\widetilde{W}^{[t+N]}\widetilde{H}^{[t+N]} - W\|= O_P\left( \frac{ \sigma p_{\max}^{\frac{1}{2}} q^{\frac{1}{2}}\max\{\sqrt{N},\sqrt{T}\}}{p_{\min}\psi_{\min}^{1/2}} \right),\\
		&(iii) \ \ \max_{t}\|\widetilde{W}^{[t+N]}\widetilde{Z}^{[t+N]\prime} - \widetilde{M}\|_F = O_P\left( \frac{\sigma p_{\max} \vartheta^{\frac{1}{2}} q^{\frac{7}{2}}\mu^{\frac{1}{2}}K \max\{\sqrt{N\log N},\sqrt{T\log T}\}}{p_{\min}^2\min\{\sqrt{N},\sqrt{T} \}} \right),\\
		&(iv) \ \ \|\widetilde{M} - M^{\star}\| = O_P\left( \frac{\sigma p_{\max}^{\frac{1}{2}} q \max\{\sqrt{N},\sqrt{ T} \}}{p_{\min}} \right),\\
		&(v) \ \ \max_{t} \|\breve{W}^{(t+N)}\breve{H}^{(t+N)} - W\|_{2,\infty} = O_P\left( \frac{\sigma p_{\max}^{\frac{1}{2}} \vartheta^{\frac{1}{2}}  q^{\frac{3}{2}}\mu^{\frac{1}{2}}K^{\frac{1}{2}}\max\{\sqrt{N\log N},\sqrt{T\log T}\}
		}{p_{\min} \psi_{\min}^{1/2} \min\{\sqrt{N},\sqrt{T} \} } \right).
	\end{align*}
\end{lemma}

\noindent \textbf{Proof of Lemma \ref{lem:CCMFY}}. First of all, by Lemmas G.1 - G.5, we have (G.1), (G.2), (G.3), (G.4) and (G.5). Hence, we have (i)-(v). Next, we prove terms in (\ref{eqb.0ada}) are $o_P(1)$. By Remark \ref{rem:simplenotation}, the first term is written as 
\begin{align}\label{eq:firstterm}
    \nonumber&\frac{1}{\sqrt{N}}\sum_{j=1}^N \omega_{jt}\varepsilon_{jt}( \widetilde{\beta}_j -\breve{\beta}^{\mathrm{full},t}_j ) 
    = N^{-\frac{1}{2}}(\widetilde{\beta} - \sqrt{N}\widetilde{W}^{[t+N]}\widetilde{H}^{[t+N]}D_{M^{\star}}^{-\frac{1}{2}})' \Omega_t \calE_t\\
    &=N^{-\frac{1}{2}}(\widetilde{\beta} - \sqrt{N}\psi_{\min}^{-1/2}\widetilde{W}^{[t+N]}H_4^{[t+N]})' \Omega_t \calE_t + \psi_{\min}^{-1/2}(H_4^{[t+N]} - \widetilde{H}^{[t+N]}D_{M^{\star}}^{-\frac{1}{2}}\psi_{\min}^{-1/2})'\widetilde{W}^{[t+N]\prime} \Omega_t \calE_t
\end{align}
where $H_4^{[N+t]}$ is a $K \times K$ matrix introduced in Claim F.2, $\Omega_t = \diag\left( \omega_{1t},\dots,\omega_{Nt} \right)$, and $\calE_t = [\varepsilon_{1t}, \dots, \varepsilon_{Nt}]'$. As noted in Claim F.2 (iii), we derive from Lemma \ref{lem:CCMFY} (iii) that 
$$
 \max_{1\leq t \leq T}\norm{\widetilde{\beta} - \sqrt{N} \psi_{\min}^{-1/2}\widetilde{W}^{[t+N]}H_4^{[t+N]}}_F = O_P\left( \frac{\sigma p_{\max} \vartheta^{\frac{1}{2}} c_{\inv}q^{\frac{9}{2}}\mu^{\frac{1}{2}}K^{\frac{3}{2}}\sqrt{N}\max\{\sqrt{N\log N},\sqrt{T\log T}\}}{p_{\min}^2 \min\{\sqrt{N},\sqrt{T} \}\psi_{\min}} \right).
$$
Hence, the first term of (\ref{eq:firstterm}) is $O_P\left( \frac{\sigma^2 p_{\max}^{\frac{3}{2}} \vartheta^{\frac{1}{2}} c_{\inv}q^{\frac{9}{2}}\mu^{\frac{1}{2}}K^{\frac{3}{2}}\sqrt{N}\max\{\sqrt{N\log N},\sqrt{T\log T}\}}{p_{\min}^2 \min\{\sqrt{N},\sqrt{T} \}\psi_{\min}} \right)$. For the second term of (\ref{eq:firstterm}), note that
\begin{align*} 
	&\max_t\|H_4^{[t+N]} - \widetilde{H}^{[t+N]} D_{M^{\star}}^{-\frac{1}{2}}\psi_{\min}^{1/2}\|\\
	& \overset{\mathrm{(i)}}{\leq} \psi_{\min}^{1/2} O_P\left( \psi_{\min}^{-1/2}\right) \left[ \max_t\|W - \widetilde{W}^{[t+N]}\widetilde{H}^{[t+N]}\| \|D_{M^{\star}}^{-\frac{1}{2}}\| + \psi_{\min}^{-1/2}\max_t\|\widetilde{W}^{[t+N]}H_4^{[t+N]} - WD_{M^{\star}}^{-\frac{1}{2}}\psi_{\min}^{1/2}\|\right] \\
	& \overset{\mathrm{(ii)}}{=}  O_P\left(\frac{\sigma p_{\max}^{\frac{1}{2}} c_{\inv} q^2 K^{\frac{1}{2}}\max\{\sqrt{N},\sqrt{ T} \}}{p_{\min} \psi_{\min}} \right).
\end{align*}
Here, (i) comes from Claim F.5 (i), and (ii) comes from Lemma \ref{lem:CCMFY} (ii) and Claim F.5 (ii). In addition, 
\begin{align*}
\max_t\|\widetilde{W}^{[t+N]\prime} \Omega_t \calE_t\| \leq \max_t\|(\widetilde{W}^{[t+N]}\widetilde{H}^{[t+N]})' \Omega_t \calE_t\| \leq \max_t\|\widetilde{W}^{[t+N]}\widetilde{H}^{[t+N]} - W\| \|\Omega_t \calE_t\| + \max_t\|W' \Omega_t \calE_t\|.
\end{align*}
From Lemma \ref{lem:CCMFY} (ii), we know $\max_t\|\widetilde{W}^{[t+N]}\widetilde{H}^{[t+N]} - W\| \|\Omega_t \calE_t\| = O_P\left( \frac{\sigma^2 p_{\max} q^{\frac{1}{2}}\sqrt{N}\max\{\sqrt{N},\sqrt{T}\}}{p_{\min}\psi_{\min}^{1/2}} \right)$. In addition, we have $\max_t\|W' \Omega_t \calE_t\| = O_P(\sigma q^{\frac{1}{2}}K^{\frac{1}{2}}\sqrt{\log T} \psi_{\min}^{1/2})$ from the matrix Bernstein inequality because $W=U_{M^{\star}}D_{M^{\star}}^{\frac{1}{2}}$. Hence, the second term of (\ref{eq:firstterm}) is 
\begin{align*}
O_P\left( \frac{\sigma^3 p_{\max}^{\frac{3}{2}} c_{\inv} q^{\frac{5}{2}} K^{\frac{1}{2}}\sqrt{N}\max\{N,T \}}{p_{\min}^2\psi_{\min}^2}
+\frac{\sigma^2 p_{\max}^{\frac{1}{2}} c_{\inv} q^{\frac{5}{2}} K \sqrt{\log T}\max\{\sqrt{N},\sqrt{T} \}}{p_{\min}\psi_{\min}}
\right).
\end{align*}
Moreover, the second term of (\ref{eqb.0ada}) can be written as 
$$
\frac{1}{\sqrt{N}}\sum_{j=1}^N \omega_{jt}\varepsilon_{jt}( \breve{\beta}^{\mathrm{full},t}_j-    \breve{\beta}^{(-t)}_j)= D_{M^{\star}}^{-\frac{1}{2}}\left(\widetilde{W}^{[t+N]}\widetilde{H}^{[t+N]} - \breve{W}^{(t+N)}\breve{H}^{(t+N)} \right)'\Omega_t \calE_t.
$$
Then, we have from Lemma \ref{lem:CCMFY} (i) that
\begin{align*}
\max_t \|D_{M^{\star}}^{-\frac{1}{2}}\left(\widetilde{W}^{[t+N]}\widetilde{H}^{[t+N]} - \breve{W}^{(t+N)}\breve{H}^{(t+N)} \right)'\Omega_t \calE_t\| 
=  O_P\left( \frac{\sigma^2 p_{\max} \vartheta^{\frac{1}{2}} q^{\frac{3}{2}}\mu^{\frac{1}{2}}K^{\frac{1}{2}}\sqrt{N}\max\{\sqrt{N\log N},\sqrt{T\log T}\}}{p_{\min} \psi_{\min} \min\{\sqrt{N},\sqrt{T} \} }\right).
\end{align*}
This completes the proof. $\square$
\bigskip
 
In addition, the following lemma shows the part in which the proofs are different depending on how we define the stopping point.

\begin{lemma} \label{lem:differentproofpoint}
Suppose assumptions of Theorem \ref{thm:generalfactor_groupclt} hold.\footnote{By Lemma F.1, it is enough to consider the assumptions of Theorem \ref{thm:generalfactor_groupclt}.} Then, we have
\begin{align*}
(1)&\ \ \max_{t} \| \frac{1}{\sqrt{N}} \sum_{j=1}^N \omega_{jt}\varepsilon_{jt}(  \breve{\beta}^{(-t)}_j - H_1'\beta_j)  \|
= O_P\left( \frac{\sigma^2 p_{\max}^{\frac{1}{2}} \vartheta^{\frac{1}{2}} q^{\frac{1}{2}}K^{\frac{1}{2}}\sqrt{\log T}\max\{\sqrt{N},\sqrt{T}\}}{p_{\min} \psi_{\min}} \right) =o_P(1),\\
(2)&\ \ \max_{t} \| \frac{1}{\sqrt{N}} \sum_{j=1}^N (\omega_{jt}-p_j)H_1'\beta_j(  \breve{\beta}^{(-t)}_j - H_1'\beta_j)  \| 
= O_P\left( \frac{ \sigma p_{\max} \vartheta q^{\frac{1}{2}}\mu^{\frac{1}{2}}K\sqrt{\log T}\max\{\sqrt{N},\sqrt{T}\}}{p_{\min}\psi_{\min}} \right) =o_P(1) .
 \end{align*}
\end{lemma}

\noindent \textbf{Proof of Lemma \ref{lem:differentproofpoint}}. 
(1)-i. Case of using $\tau^*_l$ as a stopping point:\\
Let $\xi_t\coloneqq \breve{\beta}^{(-t)}  - \beta H_1= \sqrt{N}\breve{W}^{(t+N)}\breve{H}^{(t+N)} D_{M^{\star}}^{-\frac{1}{2}}- \beta H_1$. To employ matrix Bernstein inequality, we first estimate $\max_{t}\|\xi_t\|_{2,\infty}$. Note $\|\xi_t\|_{2,\infty}\leq \sqrt{N}\psi_{\min}^{-1/2} \|\breve{W}^{(t+N)}\breve{H}^{(t+N)} - W\|_{2,\infty}$. So, by Lemma \ref{lem:CCMFY} (v), we have $\max_{t}\|\xi_t\|_{2,\infty} = O_P\left( \frac{\sigma p_{\max}^{\frac{1}{2}}\vartheta^{\frac{1}{2}} q^{\frac{3}{2}}\mu^{\frac{1}{2}}K^{\frac{1}{2}}\sqrt{N}\max\{\sqrt{N\log N},\sqrt{T\log T}\}}{p_{\min}\psi_{\min} \min\{\sqrt{N},\sqrt{T} \} } \right)$. Furthermore, we have
\begin{align*} 
	\max_t \|\xi_t\|_F &\leq \sqrt{N} \left(  \max_{t} \|\widetilde{W}^{[t+N]} \widetilde{H}^{[t+N]} - \breve{W}^{(t+N)}\breve{H}^{(t+N)} \|_F + \|W - \widetilde{W}^{[t+N]} \widetilde{H}^{[t+N]} \|_F \right)\|D_{M^{\star}}^{-\frac{1}{2}}\| \\
	&=O_P\left( \frac{\sigma p_{\max}^{\frac{1}{2}} q^{\frac{1}{2}}K^{\frac{1}{2}} \sqrt{N}\max\{\sqrt{N},\sqrt{T}\}}{p_{\min}\psi_{\min}} \right).
\end{align*}
Because $\xi_t$ only depends on $M^{\star}$ and $Y$ excluding the $t$th column of $Y$, conditioning on $\{\calM,\Omega\}$, $\{\varepsilon_{jt}\}_{j \leq N}$ are independent of $\xi_t$. Hence, $\bbE\left[ \varepsilon_{jt} | \calM, \Omega, \xi_t \right] = \bbE\left[ \varepsilon_{jt} | \calM,\Omega \right] = 0$ and, conditioning on $\{\calM,\Omega,\xi_t\}$, $\{\varepsilon_{jt}\}_{j\leq N}$ are independent across $j$. Then, by matrix Bernstein inequality, we have 
$$ \| \xi_t' \Omega_t \calE_t \|=\|\sum_{j=1}^{N} \omega_{jt} \varepsilon_{jt} \xi_{t,j}^{\prime}\| 
\leq C  \left( \sigma \log T \log N \max_{t}\|\xi_t\|_{2,\infty} + \sigma \sqrt{\log T}\max_{t}\|\xi_t\|_{F}\right)$$ 
with probability exceeding $1-O(T^{-100})$ and so, 
$\max_{t} \| \xi_t' \Omega_t \calE_t \|
= O_P\left( \frac{\sigma^2 p_{\max}^{\frac{1}{2}} \vartheta^{\frac{1}{2}} q^{\frac{1}{2}}K^{\frac{1}{2}}\sqrt{N\log T}\max\{\sqrt{N},\sqrt{T}\}}{p_{\min} \psi_{\min}} \right)$. \\
(1)-ii. Case of using $\tau^*$ as a stopping point:\\
In this case, we note that $\xi_t$ is no longer independent of $\{\varepsilon_{jt}\}_{j \leq N}$ conditioning on $\{\calM, \Omega\}$, due to the fact that $\tau^*$ does depend on the full sample. Therefore, we cannot directly apply the Bernstein inequality as in the $\tau_l^*$ case. Instead, we apply Lemma G.10 and obtain the same bound for $\max_{t} \| \xi_t' \Omega_t \calE_t \|$.\\
(2)-i. Case of using $\tau^*_l$ as a stopping point:\\
The proof is similar to that in (1-i). So, we omit it.\\
(2)-ii. Case of using $\tau^*$ as a stopping point:\\
The proof is the same as that in (1-ii) although we use Lemma G.11 instead.
$\square$

\subsection{Proof of Proposition \ref{pro:estimatef}} \label{subsec:proofofpropositionf}

First of all, by Claim F.1 (i), we can know that there is a $K\times K$ matrix $H_1$ such that $\frac{1}{\sqrt{N}}\beta H_1$ is the left singular vector of $M^{\star}$. That is, $\frac{1}{\sqrt{N}}\beta H_1=U_{M^{\star}}$. Let 
$\widetilde{B}_t \coloneqq \frac{1}{N} \sum_{j=1}^{N} \omega_{jt}\widetilde{\beta}_{j}\widetilde{\beta}_{j}^{\prime}$, $B_t^* \coloneqq \frac{1}{N} \sum_{j=1}^{N} \omega_{jt} H_1^{\prime}\beta_{j}\beta_{j}^{\prime} H_1$ and 
$B \coloneqq \frac{1}{N} \sum_{j=1}^{N} p_j H_1^{\prime}\beta_{j}\beta_{j}^{\prime} H_1$. Then, we define $H_2 \coloneqq \left( I_K + \varphi \right)H_1^{-1}$ where $\varphi \coloneqq \frac{1}{N}B^{-1}H_1^{\prime}\beta ^{\prime}\Pi\left( \beta H_1 - \widetilde{\beta} \right)$. Note that both $B$ and $H_2$ do not depend on $i$ or $t$. Because $\widehat{F}_t = \left( \sum_{j=1}^{N}\omega_{jt}\widetilde{\beta}_{j} \widetilde{\beta}_{j}^{\prime} \right)^{-1} \sum_{j=1}^{N}\omega_{jt} \widetilde{\beta}_{j} y_{jt} $ by definition, basic algebras shows the following identity:
\begin{align*} 
	&\widehat{F}_t - H_2 F_t = H_2\left( \sum_{j=1}^{N}\omega_{jt}\beta_{j} \beta_{j}^{\prime} \right)^{-1}\left( \sum_{j=1}^{N}\omega_{jt} \beta_{j} \varepsilon_{jt} \right)
	+ \sum_{d=1}^{6}\Delta_{d,t},\\
	& \Delta_{1,t} \coloneqq \widetilde{B}_t ^{-1}\frac{1}{N} \sum_{j=1}^{N}\omega_{jt}\varepsilon_{jt}\left( \widetilde{\beta}_j - H_1^{\prime}\beta_{j} \right)
	-B^{-1}H_1^{\prime}\frac{1}{N} \sum_{j=1}^{N}\left( \omega_{jt} - p_j \right) \beta_{j} F_t^{\prime} H_1^{\prime -1}\left( \widetilde{\beta}_j - H_1^{\prime}\beta_{j} \right),\\ 
	& \Delta_{2,t} \coloneqq \left(\widetilde{B}_t ^{-1} - B^{-1} \right)\frac{1}{N} \sum_{j=1}^{N}\omega_{jt}\widetilde{\beta}_j \left( \beta_{j}^{\prime}H_1 - \widetilde{\beta}_j^{\prime} \right)H_1^{-1}F_t,\\
	& \Delta_{3,t} \coloneqq B^{-1}\frac{1}{N} \sum_{j=1}^{N}\omega_{jt}\left( \widetilde{\beta}_j - H_1^{\prime}\beta_{j} \right) \left( \beta_{j}^{\prime}H_1 - \widetilde{\beta}_j^{\prime} \right)H_1^{-1}F_t,\\
	& \Delta_{4,t} \coloneqq \left(\widetilde{B}_t ^{-1} -  B_t^{*-1} \right)H_1^{\prime}\frac{1}{N} \sum_{j=1}^{N}\omega_{jt}\beta_{j}\varepsilon_{jt},\ \ \Delta_{5,t} \coloneqq \left( H_1^{-1} - H_2  \right)\left( \sum_{j=1}^{N}\omega_{jt}\beta_{j} \beta_{j}^{\prime} \right)^{-1}\left( \sum_{j=1}^{N}\omega_{jt} \beta_{j} \varepsilon_{jt} \right),\\
	& \Delta_{6,t} \coloneqq  \widetilde{B}_t ^{-1} \frac{1}{N} \sum_{j=1}^{N}\omega_{jt}\widetilde{\beta}_j M^{R}_{jt}.
\end{align*}
\noindent\textbf{Step 1.} We start from the first term of $\Delta_{1,t}$: $P_{1} \coloneqq \widetilde{B}_t ^{-1}\frac{1}{N} \sum_{j=1}^{N}\omega_{jt}\varepsilon_{jt}\left( \widetilde{\beta}_j - H_1^{\prime}\beta_{j} \right) $. We have $P_{1} = P_{1,1} + P_{1,2}$ where
$$P_{1,1} \coloneqq \widetilde{B}_t ^{-1}\frac{1}{N} \sum_{j=1}^{N}\omega_{jt}\varepsilon_{jt}\left( \widetilde{\beta}_j - \breve{\beta}_{j}^{(-t)} \right) =
\frac{1}{N}\widetilde{B}_t ^{-1}\left( \widetilde{\beta} - \sqrt{N}\breve{W}^{(N+t)}\breve{H}^{(N+t)}D_{M^{\star}}^{-\frac{1}{2}} \right)^{\prime} \Omega_t \calE_t,$$ 
$$P_{1,2} \coloneqq \widetilde{B}_t ^{-1}\frac{1}{N} \sum_{j=1}^{N}\omega_{jt}\varepsilon_{jt}\left( \breve{\beta}_{j}^{(-t)} - H_1^{\prime}\beta_{j} \right)  =
\frac{1}{N}\widetilde{B}_t ^{-1}\left( \sqrt{N}\breve{W}^{(N+t)}\breve{H}^{(N+t)}D_{M^{\star}}^{-\frac{1}{2}} - \beta  H_1 \right)^{\prime} \Omega_t \calE_t.$$ 
Note that $\max_t\|\widetilde{B}_t^{-1}\| = O_P(\frac{1}{p_{\min}})$ by Claim F.4 (iii). Hence, we have by Lemma \ref{lem:CCMFY},
\begin{align*}
&\max_t\|P_{1,1}\| \leq \max_t\|\widetilde{B}_t^{-1}\|N^{-\frac{1}{2}}\max_t\|N^{-\frac{1}{2}}( \widetilde{\beta} - \sqrt{N}\breve{W}^{(N+t)}\breve{H}^{(N+t)}D_{M^{\star}}^{-\frac{1}{2}})^{\prime} \Omega_t \calE_t\| \\
&=O_P\left( \frac{\sigma^2 p_{\max}^{\frac{3}{2}} \vartheta^{\frac{1}{2}} c_{\inv}q^{\frac{9}{2}}\mu^{\frac{1}{2}}K^{\frac{3}{2}}\max\{\sqrt{N\log N},\sqrt{T\log T}\}}{p_{\min}^3 \min\{\sqrt{N},\sqrt{T} \}\psi_{\min}}
+ \frac{\sigma^3 p_{\max}^{\frac{3}{2}} c_{\inv} q^{\frac{5}{2}} K^{\frac{1}{2}}\max\{N,T \}}{p_{\min}^3 \psi_{\min}^2}\right).
\end{align*}
Note that $\max_{t}\|P_{1,2}\| \leq \frac{1}{N}\|\widetilde{B}_t^{-1}\|\max_{t}\| \xi_t' \Omega_t \calE_t \|$. Then, using Lemma \ref{lem:differentproofpoint}, we have 
\[ 
	\max_{t}\|P_{1,2}\| = O_P\left( \frac{ \sigma^2 p_{\max}^{\frac{1}{2}}\vartheta^{\frac{1}{2}} q^{\frac{1}{2}}K^{\frac{1}{2}}\sqrt{\log T}\max\{\sqrt{N},\sqrt{T}\}}{p_{\min}^2\sqrt{N}\psi_{\min}} \right)  .
\]
\noindent\textbf{Step 2.} By using the same logic in Step 1, we can bound the second term of $\Delta_{1,t}$,\\ $P_2 \coloneqq B^{-1}H_1^{\prime}\frac{1}{N} \sum_{j=1}^{N}\left( \omega_{jt} - p_j \right) \beta_{j} F_t^{\prime} H_1^{\prime -1}\left( \widetilde{\beta}_j - H_1^{\prime}\beta_{j} \right)$ similarly. The only difference is the part using the matrix Bernstein inequality since $\{\omega_{jt}\}_{j\leq N}$ are dependent across $j$ while $\{\varepsilon_{jt}\}_{j\leq N}$ are independent across $j$. We split $P_{2}$ like $P_{2} = P_{2,1} + P_{2,2}$ where
$$
P_{2,1} \coloneqq \frac{1}{N} B^{-1}H_1^{\prime}\beta ^{\prime}\left( \Omega_t - \Pi \right)\left(\widetilde{\beta} -  \sqrt{N}\breve{W}^{(t+N)}\breve{H}^{(t+N)} D_{M^{\star}}^{-\frac{1}{2}} \right)H_1^{-1}F_t,$$
$$
P_{2,2} \coloneqq \frac{1}{N} B^{-1}H_1^{\prime}\beta ^{\prime}\left( \Omega_t - \Pi \right)\left( \sqrt{N}\breve{W}^{(t+N)}\breve{H}^{(t+N)} D_{M^{\star}}^{-\frac{1}{2}}- \beta H_1 \right)H_1^{-1}F_t.$$
By the same token as the part $P_{1,1}$ in Step 1 with the aids of Claims F.1 - F.5, we can show that 
\begin{align*}
P_{2,1} = O_P \left( \frac{\sigma^2 p_{\max}^{\frac{3}{2}}c_{\inv} q^{\frac{7}{2}}\mu K \max\{\sqrt{N}, \sqrt{T} \}}{\psi_{\min} p_{\min}^{3} \min\{\sqrt{N}, \sqrt{T} \}}
+  \frac{\sigma p_{\max}^{\frac{3}{2}} \vartheta q^{\frac{11}{2}}\mu^{\frac{3}{2}} K^{\frac{5}{2}} \max\{\sqrt{\log N}, \sqrt{\log T} \}}{ p_{\min}^{3} \min\{N, T \}}
\right).
\end{align*}
and so, we omit the proof. In addition, using Lemma \ref{lem:differentproofpoint}, the part $P_{2,2}$ can be bounded like
\[ 
	\max_{t}\|P_{2,2}\| \leq \frac{1}{\sqrt{N}}\|B^{-1}\|\max_{t}\| \frac{1}{\sqrt{N}}H_1^{\prime}\beta ^{\prime}\left( \Omega_t - \Pi \right)\xi_t \|\max_{t}\|H_1^{-1}F_t\|= O_P\left( \frac{ \sigma p_{\max} \vartheta q^{\frac{3}{2}}\mu K^{\frac{3}{2}} \sqrt{\log T}}{p_{\min}^2 \sqrt{N} \min\{\sqrt{N},\sqrt{T}\}} \right).
\]
  
\noindent\textbf{Step 3.} We bound $\max_{t}\|\Delta_{2,t}\|$. By Claim F.1 (iv), Claim F.3 (ii)
\begin{align*} 
	\max_{t}\|\Delta_{2,t}\|&
	\leq O_P(1) \max_{t}\|\widetilde{B}_t ^{-1} - B^{-1} \|\max_j\|H_1\beta_j\|p_{\max}^{\frac{1}{2}}\frac{1}{\sqrt{N}} \|\beta H_1 - \widetilde{\beta} \|_F\max_{t}\|H_1^{-1}F_t\|\\
	& =O_P \left( \frac{\sigma^2 p_{\max}^{\frac{5}{2}}c_{\inv}^2 q^5 \mu K^2 \max \{\sqrt{N}, \sqrt{T} \}}{p_{\min}^4 \min \{N, T \}\psi_{\min}} + \frac{\sigma p_{\max}^{\frac{3}{2}}c_{\inv} \vartheta q^3 \mu^{\frac{3}{2}}K^{\frac{5}{2}}\sqrt{\log T}}{\sqrt{N} \min {\sqrt{N},\sqrt{T}}}
	\right).
\end{align*}
\noindent\textbf{Step 4.} We now bound $\max_{t}\|\Delta_{3,t}\|$. By Claim F.1 (iv) and Claim F.3 (ii), we have
\begin{align*} 
	\max_{t}\|\Delta_{3,t}\|
	&\leq O_P(1) \|B^{-1}\|\frac{1}{\sqrt{N}}\|\widetilde{\beta} - \beta H_1\|\|\Pi\|\frac{1}{\sqrt{N}}\|\widetilde{\beta} - \beta H_1\|\max_{t}\|H_1^{-1}F_t\|\\
	&=O_P\left( \frac{\sigma^2 p_{\max}^2 c_{\inv}^2 q^{5} \mu^{\frac{1}{2}} K^{\frac{3}{2}}\max\{\sqrt{N},\sqrt{T} \}}{p_{\min}^3 \min\{\sqrt{N},\sqrt{T} \}\psi_{\min}}\right).
\end{align*}
\noindent\textbf{Step 5.} We estimate $\max_t\|\Delta_{4,t}\|$. By Claims F.4 (iv) and F.6 (i), we have
\begin{align*} 
	&\max_t\|\Delta_{4,t}\| \leq \frac{1}{N}\max_{t}\|\widetilde{B}_t ^{-1} - B_t^{*-1} \|\max_t\|\left( \beta H_1 \right)^{\prime}\Omega_t \calE_t\|=O_P\left( \frac{\sigma^2 p_{\max}^2 \vartheta c_{\inv} q^2 K \sqrt{\log T} \max \{\sqrt{N}, \sqrt{T} \}}{p_{\min}^3 \sqrt{N} \psi_{\min}} \right).
\end{align*}
\noindent\textbf{Step 6.} We bound $\max_{t}\|\Delta_{5,t}\|$. First, note that $H_2 -H_1^{-1} = \varphi H_1^{-1}$ and $\|\varphi\|= O_P\left( \frac{\sigma p_{\max}^{\frac{1}{2}} c_{\inv}q^{2}K^{\frac{1}{2}}\max\{\sqrt{N},\sqrt{T} \}}{p_{\min}\psi_{\min}} \right)$ as noted in the proof of Claim F.3. Moreover, by Claim F.4 (iv), we have 
$\max_t\|H_1^{-1}( \sum_{j=1}^{N}\omega_{jt} \beta_{j} \beta_{j}^{\prime} )^{-1}H_1^{\prime-1}\| = \|(NB_t^*)^{-1}\| = O_P(\frac{1}{p_{\min} N})$. Hence, by Claim F.6 (i),
\begin{align*} 
	&\max_{t}\|\Delta_{5,t}\|
	\leq \|\varphi\| \|H_1^{-1}( \sum_{j=1}^{N}\omega_{jt} \beta_{j} \beta_{j}^{\prime} )^{-1}H_1^{\prime-1}\|\max_{t}\|\left( \beta H_1 \right)^{\prime}\Omega_t \calE_t\| = O_P\left( \frac{\sigma^2 p_{\max}^{\frac{1}{2}} c_{\inv}q^{2}K^{\frac{1}{2}}\max\{\sqrt{N},\sqrt{T} \}}{p_{\min}^2\psi_{\min}} \right).
\end{align*}
\noindent\textbf{Step 7.} Lastly, we bound $\max_{t}\|\Delta_{6,t}\|$. Note that
\begin{align*}          
	\Delta_{6,t}
	&=  \left(\widetilde{B}_t ^{-1} - B^{-1} \right)\frac{1}{N} \sum_{j=1}^{N}\omega_{jt}H_1^{\prime}\beta_{j} M^{R}_{jt} +  B^{-1}\frac{1}{N} \sum_{j=1}^{N}\omega_{jt}\left(\widetilde{\beta}_j- H_1^{\prime}\beta_{j}\right)  M^{R}_{jt}\\
	& \quad + \left(\widetilde{B}_t ^{-1} - B^{-1} \right)\frac{1}{N} \sum_{j=1}^{N}\omega_{jt}\left(\widetilde{\beta}_j- H_1^{\prime}\beta_{j}\right)  M^{R}_{jt}
	+ B^{-1}\frac{1}{N} \sum_{j=1}^{N}\omega_{jt}H_1^{\prime}\beta_{j}M^{R}_{jt}.
\end{align*}
By Claims F.1, F.3 and F.4, the last term dominates the first three terms. The last term is
\[         
	\max_{t}\|B^{-1}\frac{1}{N} \sum_{j=1}^{N}\omega_{jt}H_1^{\prime}\beta_{j}M^{R}_{jt}\|
	\leq \frac{1}{\sqrt{N}}\|B^{-1}\| \|\beta H_1\|p_{\max}^{\frac{1}{2}}\max_{it}|M^{R}_{it}| =O_P\left( \frac{p_{\max}^{\frac{1}{2}}}{p_{\min}} \right)\max_{it}|M^{R}_{it}| 
\]
by Claims F.3 and F.4. This completes the proof. $\square$

\bibliographystyle{apalike}
\bibliography{reference}

\end{document}